\documentclass[fleqn,usenatbib]{mnras}

\usepackage[T1]{fontenc}

\usepackage{graphicx}	
\usepackage{amsmath}	
\usepackage{amssymb}	
\usepackage{color}

\usepackage{hyperref}
\hypersetup{
  citecolor=cyan,      
  filecolor=magenta,      
  urlcolor=magenta            
}

\usepackage{newtxtext,newtxmath}

\newcommand{\Msol}{M_{\odot}}
\newcommand{\eg}{e.g.,~}
\newcommand{\ie}{i.e.,~}
\newcommand{\cf}{cf.,~}
\newcommand{\MTOV}{M_{_{\mathrm{TOV}}}}

\title[Impact of extreme spins and mass ratios on high-mass binary
  neutron stars] {Impact of extreme spins and mass ratios on the
  post-merger observables of high-mass binary neutron stars}

\author[L. J. Papenfort et al.]{
L. Jens Papenfort,$^{1}$\thanks{E-mail: papenfort@th.physik.uni-frankfurt.de}
Elias R. Most,$^{2,3,4}$
Samuel Tootle$^{1}$
and Luciano Rezzolla$^{1,5,6}$
\\
$^{1}$Institut f\"ur Theoretische Physik, Goethe Universit\"at, Max-von-Laue-Str. 1, 60438 Frankfurt am Main, Germany\\
$^{2}$Princeton Center for Theoretical Science, Princeton University, Princeton, NJ 08544, USA\\
$^{3}$Princeton Gravity Initiative, Princeton University, Princeton, NJ 08544, USA\\
$^{4}$School of Natural Sciences, Institute for Advanced Study, Princeton, NJ 08540, USA\\
$^{5}$School of Mathematics, Trinity College, Dublin 2, Ireland\\
$^{6}$Frankfurt Institute for Advanced Studies, Ruth-Moufang-Str. 1, 60438 Frankfurt am Main, Germany\\
}

\date{Accepted XXX. Received YYY; in original form ZZZ}

\pubyear{2021}

\begin{document}
\label{firstpage}
\pagerange{\pageref{firstpage}--\pageref{lastpage}}
\maketitle

\begin{abstract}
  The gravitational-wave events GW170817 and GW190425 have led to a
  number of important insights on the equation of state of dense matter
  and the properties of neutron stars, such as their radii and the
  maximum mass. Some of these conclusions have been drawn on the basis of
  numerical-relativity simulations of binary neutron-star mergers with
  vanishing initial spins. While this may be a reasonable assumption in
  equal-mass systems, it may be violated in the presence of large mass
  asymmetries accompanied by the presence of high spins. To quantify the
  impact of high spins on multi-messenger gravitational-wave events, we
  have carried out a series of high-mass binary neutron-star mergers with
  a highly spinning primary star and large mass asymmetries that have
  been modelled self-consistently using two temperature-dependent
  equations of state. We show that, when compared with equal-mass,
  irrotational binaries, these systems can lead to significant
  differences in the remnant lifetime, in the dynamical ejecta, in the
  remnant disc masses, in the secular ejecta, and on the bulk kilonova
  properties. These differences could be exploited to remove the
  degeneracy between low- and high-spin priors in the detection of
  gravitational waves from binary neutron-star mergers.
\end{abstract}




\section{Introduction}
\label{sec:intro}
The first gravitational-wave (GW) detection of binary neutron-star (BNS)
mergers by means of the events GW170817 \citep{Abbott2017} GW190425
\citep{Abbott2020} and the potential black hole neutron star (BHNS)
binary GW190814 \citep{Abbott2020b}, have lead to a multitude of
conclusions on the nature of the most extremely compact states of
matter. Especially GW170817, thanks to its the electromagnetic
counterpart, has enabled us to put constraints on the equation of state
(EOS) of nuclear matter \citep[see \eg][]{Margalit2017, Bauswein2017b,
  Rezzolla2017, Ruiz2017, Annala2017, Radice2017b, Most2018, De2018,
  Abbott2018b, Montana2018, Raithel2018, Tews2018, Malik2018,
  Koeppel2019, Shibata2019, Nathanail2021}. The assumption that the
remnant collapsed to a black hole has resulted in an upper bound on the
maximum mass of a neutron star (NS) of $\MTOV \lesssim 2.3\,\Msol$ when
applying conservative assumptions on the merger remnant and its
properties \citep[see \eg][]{Margalit2017, Rezzolla2017, Ruiz2017,
  Shibata2019, Nathanail2021}. While the total mass of GW170817 was
presumably not large enough for a prompt collapse to a black hole (BH) at
merger, this is likely to be the case for GW190425. Given the degeneracy
between effects on the inspiral waveform induced by the tidal
deformations and the spins \citep{Hannam2013b, Favata2014, Agathos2015,
  Harry2018, ZhuX2018}, the mass ratio between the constituents of
GW170817 and GW190425 is not well constrained, leading to values of $1
\le q^{-1} \lesssim 2.5$ depending on whether low- or high-spin priors
are applied (\citealt{Abbott_etal_2018a,Abbott2020}, see also
\citet{Most2020c}). It should be noted that the spin distributions in the
case of low-spin priors are compatible with a spin that is essentially
zero and provide a tighter constraint on the mass ratio, while the
high-spin priors include extreme rotational states and yield a larger
uncertainty in the mass ratio. In particular, it has been shown that, if
the secondary of the GW190814 event was once a highly spinning NS, it
would lead to a potential rotation frequency of $f=\Omega/{2\pi}$ of
$1.21\,(1.14)\, \rm kHz$ with $\chi_{_{2}}=0.49$, using a typical NS
radius of $R = 12.5 \, \left(13\right) \, \rm{km}$ \citep{Most2020d}.

At the same time, pulsar observations have lead to a rich catalogue of
observable NSs through independent electromagnetic channels
\citep{Manchester2005-ATNF-Pulsar-Catalog, lorimer:2008, Lynch2012,
  Benacquista2013, Alsing2017, Tauris2017, Ridolfi2021}, including
observational lower limits on $\MTOV$
\citep{Demorest2010,Antoniadis2013,Fonseca2016,Arzoumanian2018,Cromartie2019},
the evidence for extreme rotational frequencies \citep{Hessels2006}, as
well as binary-pulsar systems with significant mass asymmetries
\citep{Martinez2015,Lazarus2016,Tauris2019}, companions with appreciable
spin frequencies \citep{Lyne04,Stovall2018}, and masses in X-ray binaries
as low as $\approx 1.1 \, \Msol$ \citep{Rawls2011}. Based on these
observational evidences, we hereafter assume a conservative mass range
for the mass of a nonrotating NS to be $1.1 \lesssim M_{_{\rm NS}}
\lesssim 2.3$ for NSs in BNS systems. In addition, the increasingly
sophisticated parametric studies on population synthesis and analyses of
possible (sub-)population binary-formation channels \citep[see
  \eg][]{Dominik2013, Tauris2017, Kruckow2018, Andrews2019}, suggest that
the ratio in masses in a BNS system should lie in the range $0.5 \lesssim
q \le 1.0$.

On the other hand, the range of spins in BNS systems is far less
constrained. The highly spinning millisecond pulsars known today are most
probably all recycled pulsars which gained angular momentum through
different accretion processes in binaries involving a donor star
\citep{Burderi1999, Tauris2012, Miller2014, Tauris2015, Tauris2017},
where the most mass is accreted in common-envelope evolutions.
Theoretical estimations of the amount of spin-up by mass accretion
processes show that significant mass fractions have to be accreted to
reach the sub-millisecond scale \citep{Burderi1999,Tauris2012}, while
simulations of the common-envelope phase suggest a limit of $M_{\rm acc}
\lesssim 0.1 \, \Msol$ \citep{MacLeod2015b,Cruz2020b} with limited
efficiency of angular momentum transport
\citep{Murguia-Berthier2017845}. Nonetheless, there is evidence of
systems potentially accreting with rates far beyond the Eddington limit
\citep{Israel2017}, while the magnetic field should be damped quickly and
thus the star would not be visible as a pulsar on longer time-scales

Over the past two decades, a large number of theoretical studies have
been carried out to understand the effects of mass asymmetry in BNS
mergers, mostly considering rather small asymmetries, \ie $q \gtrsim 0.7$
with irrotational NS constituents \citep[see, \eg][]{Shibata:2003ga,
  Rezzolla:2010, Bauswein2013, Neilsen2014, Dietrich:2015b, Radice2016,
  Dietrich2017, Radice2017b, Papenfort2018, Most2021c}. Smaller mass
ratios, \eg $q \lesssim 0.5$, have been simulated only with a piecewise
polytropic EOS which was either stiff \citep{Dietrich:2015b,
  Dietrich2017}, or rather soft \citep{Tichy2019}. Theses studies have
then been followed by the first evolutions of BNS systems with $q\approx
0.45$ adopting tabulated temperature and electron-fraction dependent EOSs
\citep{Most2020e,Papenfort2020}. In summary, this large bulk of work
points out that a stronger mass asymmetry leads to a more violent tidal
disruption of the secondary, which is accompanied by an increase of the
dynamical ejecta through tidal tails at lower electron fraction, while
the systems retain a larger disc mass after the collapse of the remnant
as already pointed out by \citet{Shibata06a} and \citet{Rezzolla:2010}.

Additionally, attention has also been paid to investigation of binaries
with various degrees of spin prior to the merger \citep{Kastaun2013,
  Bernuzzi2013, East2016, Dietrich2017c, East2019, Most2019,
  Most2020e}. These works have pointed out that a significant amount of
stellar spin aligned with orbital angular momentum can increase the
lifetime of the merger remnant and leave an imprint on the final spin of
the BH produced when the remnant collapses, attaining an upper limit of
$J/M^2\simeq0.89$ the dimensionless spin \citep{Kastaun2013,
  Bernuzzi2013}. Interestingly, the combination of high spins of up to
$|\chi_{_{1,2}}| = 0.756$ and highly eccentric orbits can also increase
the lifetime of the remnant \citep{East2016}, while large aligned spins
alone can boost the development of a one-arm instability in the remnants
and increase the disc masses \citep{East2019}. At the same time, the
comparison of moderate spins with $|\chi_{_{1,2}}| = 0.1$ in binaries
with mass asymmetry in the range $q\in[0.66, 1]$, has shown that these
are subdominant to the effects of mass asymmetry \citep{Dietrich2017c}.
More recently, BNS configurations with strongly spinning and equal-mass
constituents having (anti-)aligned spins and fully temperature-dependent
EOSs have been shown to exhibit a strong suppression of the dynamical
ejecta for aligned spins \citep{Most2019}. In addition, it has been shown
how highly asymmetric binaries with $q=0.5$ and significant spin of the
massive primary (whose mass is close to the maximum mass for either
rotating or nonrotating models), can produce dynamical ejecta with a
significant fast component that can be used to discriminate this BNS from
a BH-NS binary \citep{Most2020e}.

At the same time, considerable effort has been put into constraining the
the threshold mass to prompt collapse at merger, $M_{_{\rm th}}$, that
is, the critical mass above which the merged object collapses on a
dynamical timescale \citep{Bauswein2013, Koeppel2019, Agathos2019,
  Bauswein2020, Tootle2021, Perego2021, Koelsch2021}.  The prompt
collapse scenario provides important constraints on follow-up
electromagnetic radiation since it affects the amount of unbound matter
in the system, with the latter being useful to constrain the survival
time of the remnant \citep{Gill2019}. It was realised early that the
threshold mass can be parametrized at least approximately in terms of the
mass of the binary and hence of the maximum mass $\MTOV$
\citep{Shibata05c,Baiotti08,Bauswein2013}, such that $M_{_{\rm th}}
\propto \MTOV$. More recent studies have explored this process including
more realistic, temperature-dependent EOSs
\citep{Koeppel2019,Agathos2019,Bauswein2020}, incorporating the effects
of mass asymmetry \citep{Bauswein2020c} and non-negligible spin in the
binary \citep{Tootle2021}. Overall, these studies have shown that smaller
mass ratios reduce the threshold mass and increase both the dynamically
ejected mass and the mass in the disc mass around the promptly formed
BH. At the same time, the inclusion of spin reveals that $M_{_{\rm th}}$
can increase (decrease) by 5\% (10\%) for binaries that have spins
aligned (antialigned) with the orbital angular momentum and that the
threshold mass has a non-monotonic dependence on the mass asymmetry in
the system \citep{Tootle2021}.

We here provide the first systematic study on BNS configurations
significantly above the irrotational threshold mass in conjunction with a
highly spinning primary companion in a binary with strong mass
asymmetries. We point out important differences between the stability of
the merger remnant, its post-merger properties, the resulting ejecta, and
the disc masses after collapse, when extreme spins are considered. We do
this after employing two tabulated temperature- and electron-fraction
dependent EOSs and a neutrino leakage scheme. Overall, we find that the
inclusion of extreme spin rates can provide significant differences in
the threshold mass, dynamical ejecta, remnant disc masses and therefore
in the amount of secular ejecta. In turn, these differences could be used
to discriminate between low- and high-spin priors in the detection of
GWs from BNS mergers with such high masses.

The plan of the paper is as follows. In Sec.~\ref{sec:methods} we present
our numerical setup, while in Sec.~\ref{sec:results} we illustrate the
dynamics from the various models considered, contrasting the similarities
and the differences, and collecting the most salient features. Finally,
Sec.~\ref{sec:discussion} provides a concluding discussion and the
prospects of future work.

\begin{table*}
\begin{centering}
\addtolength{\tabcolsep}{-2pt}    
\begin{tabular}{c|c|c|c|c|c|c|c|c|c|c|c|c|c|c|c}
 binary model & EOS & $M_{1}$  & $M_{2}$ & $M_{\rm{b},1}$ & $M_{\rm{b},2}$ & $M_{_{\infty}}$ & $M_{_{\infty}}$
  & $M_{_{\rm ADM}}$ & $q$ & $\tilde{\chi}_{_{\rm init}}$
  & $\chi_{_{1}}$  & $\chi_{_{\rm eff}}$ & $M_{_{\rm disc}}$
  & $\tau_c$ & lifetime \\

 & & $\left[M_\odot\right]$ & $\left[M_\odot\right]$ &
 $\left[M_\odot\right]$ & $\left[M_\odot\right]$ & $[M^{^{1,0}}_{\rm
     th}]$ & $\left[M_\odot\right]$ & $\left[M_\odot\right]$ & & & & & $[M_{\odot}]$ & $\left[\rm ms\right]$ & symbol \tabularnewline

\hline
\hline
    $\texttt{TNT-10.5-0.30-0.837}$ & \texttt{TNTYST}           & $1.741$ & $1.458$ & $2.155$ & $1.983$ & $1.105$ & $3.199$ & $3.166$ & $0.837$ & $0.944$ & $0.30$ & $0.158$ & $0.009$ & $  0.7$ & $\bigtriangledown$  \\
    $\texttt{TNT-10.5-0.30-1.000}$ & \texttt{TNTYST}           & $1.600$ & $1.600$ & $1.799$ & $1.809$ & $1.105$ & $3.199$ & $3.166$ & $1.000$ & $0.939$ & $0.30$ & $0.150$ & $0.019$ & $  0.6$ & $\bigtriangledown$  \\
    $\texttt{TNT-10.5-0.45-0.675}$ & \texttt{TNTYST}           & $1.910$ & $1.289$ & $2.192$ & $1.418$ & $1.105$ & $3.199$ & $3.167$ & $0.675$ & $0.978$ & $0.45$ & $0.268$ & $0.152$ & $  2.4$ & $\bigtriangledown$  \\
    $\texttt{TNT-10.5-0.45-0.837}$ & \texttt{TNTYST}           & $1.741$ & $1.458$ & $1.969$ & $1.628$ & $1.105$ & $3.199$ & $3.166$ & $0.837$ & $0.981$ & $0.45$ & $0.244$ & $0.029$ & $  1.0$ & $\bigtriangledown$  \\
    $\texttt{TNT-10.5-0.45-1.000}$ & \texttt{TNTYST}           & $1.600$ & $1.600$ & $1.788$ & $1.809$ & $1.105$ & $3.199$ & $3.166$ & $1.000$ & $0.970$ & $0.45$ & $0.225$ & $0.092$ & $  0.7$ & $\bigtriangledown$  \\
    $\texttt{TNT-10.5-0.60-0.675}$ & \texttt{TNTYST}           & $1.910$ & $1.289$ & $2.166$ & $1.418$ & $1.105$ & $3.199$ & $3.168$ & $0.675$ & $1.023$ & $0.60$ & $0.358$ & $0.227$ & $>36.4$ & $\bigtriangleup  $  \\
    $\texttt{TNT-10.5-0.60-0.837}$ & \texttt{TNTYST}           & $1.741$ & $1.458$ & $1.949$ & $1.628$ & $1.105$ & $3.199$ & $3.166$ & $0.837$ & $1.018$ & $0.60$ & $0.326$ & $0.118$ & $  5.7$ & $\square$           \\
    $\texttt{TNT-10.5-0.60-1.000}$ & \texttt{TNTYST}           & $1.600$ & $1.600$ & $1.772$ & $1.809$ & $1.105$ & $3.199$ & $3.165$ & $1.000$ & $1.001$ & $0.60$ & $0.300$ & $0.176$ & $  1.2$ & $\bigtriangledown$  \\
\hline                                                                                                                                                  
    $\texttt{TNT-05.0-0.45-0.600}$ & \texttt{TNTYST}           & $1.899$ & $1.139$ & $2.176$ & $1.237$ & $1.050$ & $3.039$ & $3.010$ & $0.600$ & $1.034$ & $0.45$ & $0.281$ & $0.240$ & $>34.2$ & $\bigtriangleup  $  \\
    $\texttt{TNT-05.0-0.45-0.800}$ & \texttt{TNTYST}           & $1.688$ & $1.350$ & $1.899$ & $1.492$ & $1.050$ & $3.039$ & $3.006$ & $0.800$ & $1.048$ & $0.45$ & $0.250$ & $0.088$ & $  9.9$ & $\square$           \\
    $\texttt{TNT-05.0-0.45-1.000}$ & \texttt{TNTYST}           & $1.519$ & $1.519$ & $1.686$ & $1.704$ & $1.050$ & $3.039$ & $3.006$ & $1.000$ & $1.036$ & $0.45$ & $0.225$ & $0.100$ & $  6.0$ & $\square$           \\
    $\texttt{TNT-05.0-0.60-0.600}$ & \texttt{TNTYST}           & $1.899$ & $1.139$ & $2.151$ & $1.237$ & $1.050$ & $3.039$ & $3.010$ & $0.600$ & $1.086$ & $0.60$ & $0.375$ & $0.262$ & $>34.8$ & $\bigtriangleup  $  \\
    $\texttt{TNT-05.0-0.60-0.800}$ & \texttt{TNTYST}           & $1.688$ & $1.350$ & $1.883$ & $1.492$ & $1.050$ & $3.039$ & $3.007$ & $0.800$ & $1.089$ & $0.60$ & $0.333$ & $0.137$ & $>35.9$ & $\bigtriangleup  $  \\
    $\texttt{TNT-05.0-0.60-1.000}$ & \texttt{TNTYST}           & $1.519$ & $1.519$ & $1.674$ & $1.704$ & $1.050$ & $3.039$ & $3.007$ & $1.000$ & $1.068$ & $0.60$ & $0.300$ & $0.231$ & $>36.2$ & $\bigtriangleup  $  \\
\hline                                                                                                                                                  
\hline                                                                                                                                                  
    $\texttt{BHB-10.8-0.30-0.837}$ & \texttt{BHB}$\Lambda\Phi$ & $1.904$ & $1.595$ & $2.155$ & $1.771$ & $1.109$ & $3.500$ & $3.460$ & $0.837$ & $0.919$ & $0.30$ & $0.163$ & $0.034$ & $  0.7$ & $\bigtriangledown$  \\
    $\texttt{BHB-10.8-0.45-0.675}$ & \texttt{BHB}$\Lambda\Phi$ & $2.089$ & $1.410$ & $2.381$ & $1.544$ & $1.109$ & $3.500$ & $3.461$ & $0.675$ & $0.954$ & $0.45$ & $0.268$ & $0.187$ & $  0.7$ & $\bigtriangledown$  \\
    $\texttt{BHB-10.8-0.45-0.837}$ & \texttt{BHB}$\Lambda\Phi$ & $1.904$ & $1.595$ & $2.140$ & $1.771$ & $1.109$ & $3.500$ & $3.460$ & $0.837$ & $0.956$ & $0.45$ & $0.244$ & $0.065$ & $  0.8$ & $\bigtriangledown$  \\
    $\texttt{BHB-10.8-0.60-0.675}$ & \texttt{BHB}$\Lambda\Phi$ & $2.089$ & $1.410$ & $2.355$ & $1.544$ & $1.109$ & $3.500$ & $3.461$ & $0.675$ & $0.998$ & $0.60$ & $0.358$ & $0.259$ & $  2.4$ & $\bigtriangledown$  \\
    $\texttt{BHB-10.8-0.60-0.837}$ & \texttt{BHB}$\Lambda\Phi$ & $1.904$ & $1.595$ & $2.121$ & $1.771$ & $1.109$ & $3.500$ & $3.460$ & $0.837$ & $0.992$ & $0.60$ & $0.326$ & $0.162$ & $  0.9$ & $\bigtriangledown$  \\
\hline                                                                                                                                                  
    $\texttt{BHB-05.0-0.60-0.600}$ & \texttt{BHB}$\Lambda\Phi$ & $2.071$ & $1.242$ & $2.331$ & $1.343$ & $1.050$ & $3.314$ & $3.279$ & $0.600$ & $1.065$ & $0.60$ & $0.375$ & $0.289$ & $  8.4$ & $\square$           \\
    $\texttt{BHB-05.0-0.60-0.800}$ & \texttt{BHB}$\Lambda\Phi$ & $1.841$ & $1.472$ & $2.042$ & $1.619$ & $1.050$ & $3.314$ & $3.277$ & $0.800$ & $1.066$ & $0.60$ & $0.333$ & $0.158$ & $>39.1$ & $\bigtriangleup  $  \\
    $\texttt{BHB-05.0-0.60-1.000}$ & \texttt{BHB}$\Lambda\Phi$ & $1.657$ & $1.657$ & $1.817$ & $1.848$ & $1.050$ & $3.314$ & $3.277$ & $1.000$ & $1.045$ & $0.60$ & $0.300$ & $0.191$ & $  0.9$ & $\bigtriangledown$  \\
\hline                                                                                                                                                  
    $\texttt{BHB-02.5-0.45-0.600}$ & \texttt{BHB}$\Lambda\Phi$ & $2.022$ & $1.213$ & $2.292$ & $1.308$ & $1.025$ & $3.235$ & $3.202$ & $0.600$ & $1.044$ & $0.45$ & $0.281$ & $0.262$ & $  1.5$ & $\bigtriangledown$  \\
    $\texttt{BHB-02.5-0.45-0.800}$ & \texttt{BHB}$\Lambda\Phi$ & $1.797$ & $1.437$ & $2.004$ & $1.576$ & $1.025$ & $3.235$ & $3.199$ & $0.800$ & $1.057$ & $0.45$ & $0.250$ & $0.146$ & $ 25.2$ & $\bigtriangleup  $  \\
    $\texttt{BHB-02.5-0.45-1.000}$ & \texttt{BHB}$\Lambda\Phi$ & $1.617$ & $1.617$ & $1.781$ & $1.798$ & $1.025$ & $3.235$ & $3.199$ & $1.000$ & $1.044$ & $0.45$ & $0.225$ & $0.128$ & $  2.9$ & $\bigtriangledown$  \\
    $\texttt{BHB-02.5-0.60-0.600}$ & \texttt{BHB}$\Lambda\Phi$ & $2.022$ & $1.213$ & $2.268$ & $1.308$ & $1.025$ & $3.235$ & $3.202$ & $0.600$ & $1.097$ & $0.60$ & $0.375$ & $0.307$ & $ 26.5$ & $\bigtriangleup$    \\
    $\texttt{BHB-02.5-0.60-0.800}$ & \texttt{BHB}$\Lambda\Phi$ & $1.797$ & $1.437$ & $1.988$ & $1.576$ & $1.025$ & $3.235$ & $3.199$ & $0.800$ & $1.098$ & $0.60$ & $0.333$ & $0.153$ & $>38.8$ & $\bigtriangleup  $  \\
    $\texttt{BHB-02.5-0.60-1.000}$ & \texttt{BHB}$\Lambda\Phi$ & $1.617$ & $1.617$ & $1.770$ & $1.798$ & $1.025$ & $3.235$ & $3.199$ & $1.000$ & $1.077$ & $0.60$ & $0.300$ & $0.246$ & $  1.3$ & $\bigtriangledown$  \\

\end{tabular}
\par\end{centering}
\caption{Summary of the binary configurations considered. The different
  columns list: the binary model, the EOS, the gravitational masses
  $M_{1,2}$ of both stars in isolation, the baryonic masses
  $M_{\rm{b},1,2}$, the ratio between the total gravitational mass at
  infinite separation $M_{_{\infty}}$ and the irrotational threshold mass
  of the given EOS $M^{^{1,0}}_{\rm th}$, the ADM mass of the initial
  data $M_{_{\rm ADM}}$, the initial rescaled total dimensionless angular
  momentum $\tilde{\chi}_{_{\rm init}}$ (see Eq.~\eqref{eqn:al}), the
  mass ratio $q := M_{2} / M_{1} \leq1$, the dimensionless spin angular
  momentum of the primary star $\chi_{_{1}}$, the resulting effective
  spin of the system $\chi_{_{\rm eff}}$, the post-merger remnant disc
  mass $M_{_{\rm disc}}$, the survival time of the remnant after merger
  $\tau_c$, and a symbol indicating the remnant lifetime. In particular,
  a short-lived remnant is marked by $\bigtriangledown$, a medium-lived
  remnant by $\square$ and a long-lived remnant by $\bigtriangleup$ (see
  Sec.~\ref{sec:results-stability} for a definition). The secondary is
  assumed to be irrotational in all cases, \ie $\chi_{_{2}} = 0$ and
  $M_{_{\infty}}$ is fixed to be larger than the corresponding threshold
  mass $M^{^{1,0}}_{\rm th}$ of the given EOS (see Table
  \ref{table:eos-properties}). }
\label{tab:initial}
\end{table*}

\section{Methods}
\label{sec:methods}

\begin{table}
  \centering
  \begin{tabular}{lccccc}
    EOS & $M_{_{\rm TOV}}$ & $R_{_{1.4}}$            & $\mathcal{C}_{1.4}$ & $M^{^{1,0}}_{\rm th}$ & $M^{^{1,0}}_{\rm th}$  \\
        & $[M_{\odot}]$ &   $[{\rm km}]$        &          & $[M_{_{\rm TOV}}]$ & $[M_{\odot}]$ \\ 
    \hline
    \hline
    \texttt{TNTYST}$+$          &           2.23 &        11.54 &   0.28 & 1.298 & 2.894 \\
    \texttt{BHB}$\Lambda\Phi$   &           2.10 &        13.22 &   0.23 & 1.503 & 3.156 \\
    \hline
  \end{tabular}
  \caption{Properties of the tabulated nuclear EOSs utilised in this
    study: the softer \texttt{TNTYST} \citep{Togashi2017} and the stiffer
    \texttt{BHB}$\Lambda\Phi$ \citep{Banik2014}. We report the maximum
    mass of a nonrotating NS, $M_{\rm TOV}$, together with the radius
    at a mass of $1.4\,M_\odot$, $R_{1.4}$, and the corresponding
    compactness, $\mathcal{C}_{1.4}:=1.4/R_{1.4}$. Also shown are the
    threshold masses as reported by \citet{Koeppel2019} for an equal-mass
    irrotational BNS, $M^{^{1,0}}_{\rm th}$.}
  \label{table:eos-properties}
\end{table}

The equations of relativistic hydrodynamics in dynamical spacetimes are
solved by the \texttt{Frankfurt\-/IllinoisGRMHD} code (\texttt{FIL})
\citep{Most2019b,Most2018b} derived from the \texttt{IllinoisGRMHD} code
\citep{Etienne2015}, and is built on top of the \texttt{Einstein Toolkit}
\citep{loeffler_2011_et}. It implements high-order (fourth) conservative
finite-difference methods \citep{DelZanna2007} for the fluid and magnetic
field evolution, supports temperature and electron-fraction dependent
EOSs, and neutrino cooling and weak interactions by a neutrino leakage
scheme \citep{Ruffert96b,Rosswog:2003b,Galeazzi2013}. The properties of
nuclear matter are modeled using the \texttt{TNTYST} \citep{Togashi2017}
and \texttt{BHB}$\Lambda\Phi$ \citep{Banik2014} EOS. The latter is based
on the DD2 EOS \citep{Hempel2010} but incorporates hyperons
\citep{Banik2014} at high densities. Characteristic properties of both
EOS can be found in Tab.~\ref{table:eos-properties}, with
\texttt{BHB}$\Lambda\Phi$ being stiffer and leading to larger radii of
the stars. Besides, these EOS cover a range of maximum masses and radii
consistent with GW170817.

\begin{figure*}
  \centering
  \centering
  \includegraphics[width=0.99\textwidth]{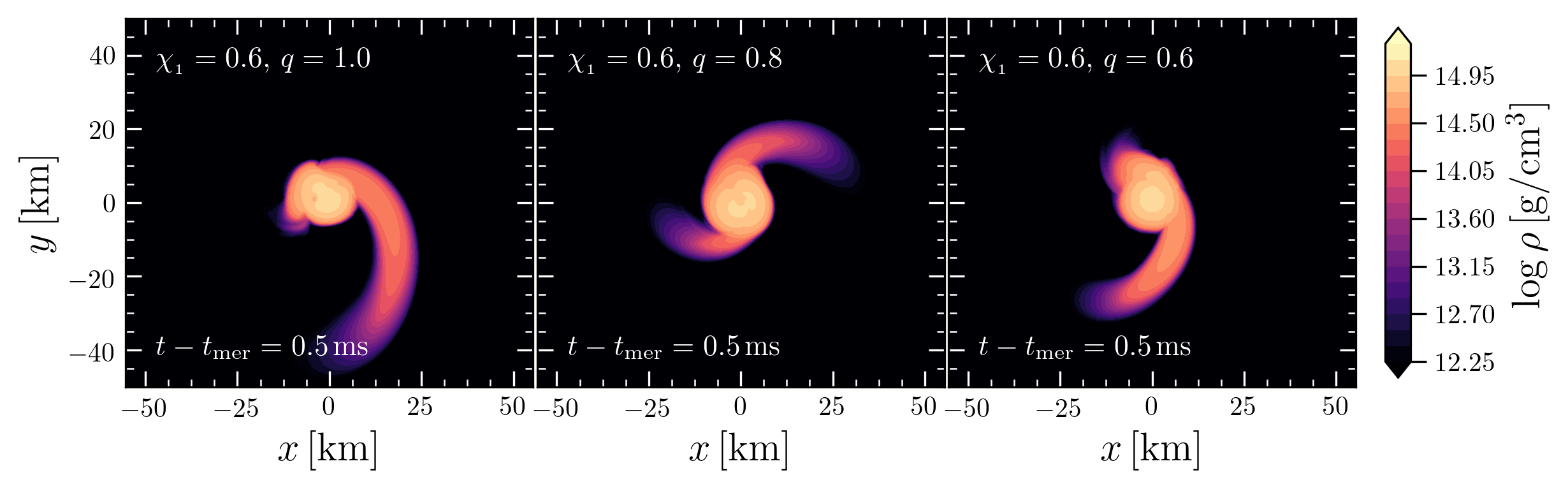}
  \caption{Rest-mass density distribution $\rho$ on the equatorial plane
    of three representative simulations shortly after the merger, \ie
    $t-t_{\rm mer} = 0.5 \, \rm{ms}$. The three binaries have
    $\chi_{_{1}} = 0.6$ and mass ratios $q \in \left\{1.0, 0.8, 0.6
    \right\}$ for the \texttt{TNTYST} at $M_{\infty}/M^{^{1,0}}_{\rm th}
    = 1.05$. At mass symmetry (\textit{left}), the highly spinning
    primary is tidally disrupted, while at strong mass asymmetry
    (\textit{right}) it is the irrotational secondary that is disrupted
    instead. At this spin magnitude, the intermediate mass ratio $q =
    0.8$ (\textit{middle}) shows merger dynamics closer to a typical
    irrotational merger.}
  \label{fig:merger-togs}
\end{figure*}

The code also implements a constraint-damping formulation of the Z4 form
of the Einstein field equations \citep{Bona:2003fj, Bernuzzi:2009ex,
  Alic:2011a}. We model the initial binary configurations via the
solution of the constraint equations solved using the extended conformal
thin-sandwich (XCTS) formalism \citep{Pfeiffer:2002iy,
  Pfeiffer:2005}. The highly spinning companion is constructed following
the formalism proposed by \citet{Tichy12}, for which the velocity field
of the star is a combination of an irrotational and of a uniformly
rotating part \citep{Tacik15, Tsokaros2015, Tsokaros2016,
  Papenfort2020}. This is implemented using the
\texttt{FUKA}\footnote{https://kadath.obspm.fr/fuka/} code
\citep{Papenfort2020}, which extends to general compact binary initial
data the \texttt{Kadath} spectral library \citep{Grandclement09}. Note
that although \texttt{FIL} solves the equations of general-relativistic
magnetohydrodynamics, for simplicity, we set the magnetic field to zero
in our analysis.

The computational domain is managed by the adaptive mesh-refinement
driver \texttt{Carpet} \citep{Schnetter-etal-03b} and features a set of
nested Cartesian box-in-box levels. We have here used total of eight
levels of refinement, with the finest one having a resolution of $h =
0.16\,M_{\odot} \simeq 236\, {\rm m}$ and with outermost box extending up
to $\simeq 6000\,\rm km$. The resolution employed is rather high and
sufficient to capture accurately the NSs, whose size depends on the mass
ratio and on the spin rate (stars rotating near the mass shedding acquire
a rather flattened shape). We note that we have conducted all of
the simulations also at a lower resolution of $h = 0.2\,M_{\odot}
\simeq 295\, {\rm m}$ and -- in some cases -- with a higher resolution
of $h = 0.133\,M_{\odot} \simeq 197\, {\rm m}$. The uncertainties that
can be measured through this comparative analysis indicate that the
errors are of a few tens of a percent at most. A more extended
discussion detailing the various uncertainties in the different
quantities can be found in the Appendix \ref{appendix_A}.

Using as a reference the threshold mass for an equal-mass irrotational
system as predicted by the quasi-universal relation of
\citet{Koeppel2019}, hereafter simply $M^{^{1,0}}_{\rm th}$, we have
explored BNSs where the binary has masses up to $\approx
1.1\,M^{^{1,0}}_{\rm th}$ and the mass ratio can be as small as $q =
0.6$\footnote{For simplicity, we have here employed a rescaling
  in terms of threshold mass for an equal-mass irrotational binary
  $M^{^{1,0}}_{\rm th}$, thus ignoring the corrections that are
  introduced by spin and mass ratio and presented by
  \citet{Tootle2021}. While these are interesting, they do not provide
  additional information on the systematic behaviour.}. This rather
extreme systems can be constructed because the primary is very highly
spinning, with a dimensionless spin in the range $\chi_{_{1}} := J_{_{\rm
    ADM, 1}} / M^2_{_{\rm ADM, 1}} \in \left\{0.30, 0.45, 0.60\right \}$,
where $J_{_{\rm ADM, 1}}$ and $M_{_{\rm ADM, 1}}$ are the ADM angular
momentum and mass of the primary in isolation, respectively. In all
cases, the spin axis is aligned with the orbital angular momentum, mostly
for convenience rather than for realism. While placing all of the spin in
the system in the primary is not only a convenient choice, it is in all
likelihood also the most realistic one. Indeed, the astrophysical
scenario reproduced with our initial data corresponds, from an
evolutionary point of view, to the case in which the primary NS has
gained the additional spin angular momentum through mass transfer from
the progenitor star of the secondary NS \citep[see \eg][]{Tauris2017}. In
this scenario, therefore, the secondary has a spin which is much smaller
than the primary and can effectively be treated as irrotational; a
partially more extended investigation of the space of parameters to
include less extreme spin magnitudes has already been presented by
\citet{Tootle2021} and will part of our future work. Finally, all
binaries start with an initial separation of $45\,{\rm km}$ and their
initial properties are given in in Tab.~\ref{tab:initial}.

\section{Results}
\label{sec:results}

In this section we present the results of our numerical evolutions and
discuss the effects introduced by the large amounts of spin in the
primary star.

In total, we conducted 14 simulations for each EOS, with a total of 28
simulations. These consist of two sets for the \texttt{TNTYST} EOS and
three for the \texttt{BHB}$\Lambda\Phi$ EOS. In each set, we select a
binary after fixing the ADM mass of the system at infinite separation,
$M_{_{\infty}}$, in terms of the threshold mass for an equal-mass
irrotational binary as found by \citet{Koeppel2019} for these EOSs; this
quantity can be taken as a measure of the ``mass criticality'' in the
system. More specifically, we consider binaries with ADM mass
$M_{_{\infty}} / M^{^{1,0}}_{\rm th} \in \left\{1.050, 1.105 \right\}$
for \texttt{TNTYST} and $M_{_{\infty}} / M^{^{1,0}}_{\rm th} \in
\left\{1.025, 1.050, 1.109 \right\}$ for \texttt{BHB}$\Lambda\Phi$.
Furthermore, for each binary we vary the mass asymmetry $q$ between the
two stars with $q \in \left[0.6, 1.0\right]$, but also the amount of spin
of the primary, $\chi_{_{1}}$, with $\chi_{_{1}} \in \left[0.3,
  0.6\right]$. In this way, we can construct a table of 28 different
configurations whose properties can be found in Tab.~\ref{tab:initial}.

In the following subsections we present the results concerning the merger
and post-merger dynamics, the stability of the remnant, as well as the
dynamical mass ejection, the remnant disc masses, which are important for
the secular ejecta and their implications on the resulting kilonova
emission.

\begin{figure*}
  \centering
  \centering
  \includegraphics[width=0.99\textwidth]{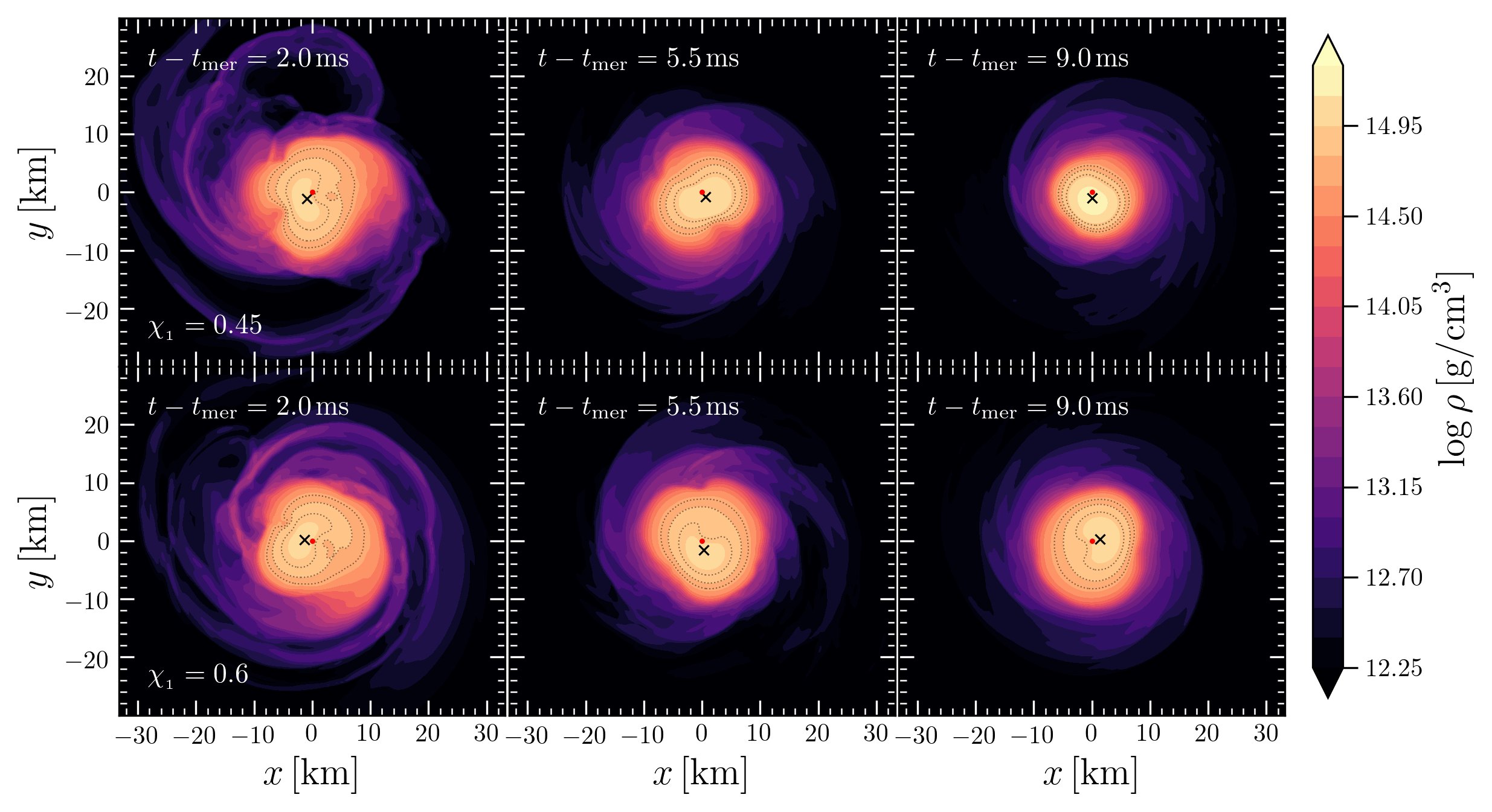}
  \caption{Rest-mass density distribution $\rho$ on the equatorial plane
    of the two models $\texttt{TNT-05.0-0.45-0.800}$ and
    $\texttt{TNT-05.0-0.60-0.800}$ at different times in the early
    post-merger phase. Both systems exhibit a total mass which is $5 \%$
    above $M^{^{1,0}}_{\rm th}$ of the \texttt{TNTYST} EOS (see
    Tab.~\ref{tab:initial}) with a dimensionless spin of the primary
    companion of $\chi_{_{1}} = 0.45$ (\textit{top}) and $\chi_{_{1}} =
    0.6$ (\textit{bottom}), respectively. The black cross denotes the
    point of highest rest-mass density and the red dot marks the origin
    of the coordinates, representing the approximate location of the axis
    of rotation. In addition, three dotted contours at $\log \rho \in
    \left\{14.7,14.8,14.9\right\}$ highlight the asymmetries of the
    matter distribution.}
  \label{fig:one-arm-tog}
\end{figure*}

\subsection{Merger and post-merger dynamics}
\label{sec:results-post-merger}

As is well-known already from post-Newtonian dynamics, binary systems
with highly spinning constituents experience longer inspirals with
increasing spin of the primary due to the spin-orbit coupling and the net
increased angular momentum, thus exhibiting different times of merger
\citep{Bernuzzi2013,Dietrich:2015b,Dietrich2017c,Most2019}. While this is
interesting in its own right and important when constructing
high-precision waveforms, hereafter we will concentrate on the violent
merger process at these high masses, spins, and mass asymmetry.

As a representative set of models, we show in Fig.~\ref{fig:merger-togs}
snapshots of the rest-mass density at $t-t_{\rm mer} = 0.5 \, \rm{ms}$
after the merger of three models at different mass ratios and with the
highest primary spin $\chi_{_{1}} = 0.6$ for the \texttt{TNTYST} EOS. For
all of them the total mass is $5 \%$ above the irrotational threshold
mass, \ie $M_{_{\infty}} / M^{^{1,0}}_{\rm th} = 1.05$.

At these high spins, there is an interesting transition in the star that
is first disrupted tidally. More specifically, in the $q=1$ equal-mass
case (left panel in Fig.~\ref{fig:merger-togs}), the disruption is
suffered by the primary, which is also the only star carrying angular
momentum. However, when a large asymmetry in the mass is present, as for
the $q=0.6$ case (right panel in Fig.~\ref{fig:merger-togs}), then the
disruption is suffered by the secondary, which is less massive and
irrotational. In the intermediate case of $q = 0.8$ (middle panel in
Fig.~\ref{fig:merger-togs}), on the other hand, the disruption follows a
pattern already encountered in the case of irrotational unequal-mass
binaries. More specifically, the secondary appears to be disrupted first
and accretes mass and angular momentum on the primary, that is also
disrupted as a consequence. A more detailed investigation of this
scenario, possibly including tracer particles \citep{Bovard2016}, will
help determine with more precision the fate of the two star in this mass
ratio, see also \citet{Most2020e} for a similar scenario.

For binaries with the same spin magnitude but with larger mass
criticality, \ie $M_{_{\infty}} / M^{^{1,0}}_{\rm th}=1.105$, the
primary is less affected at merger in the case of $q = 1$ and the tidal
disruption much more contained. This trend becomes even stronger for
equal-mass binaries with smaller primary spins, \ie $\chi_{_{1}} = 0.45$,
where the dynamics is very similar to that of irrotational
binaries. Nonetheless, the spin still has a noticeable impact at merger,
especially in the $q = 0.6$ cases, where the secondary is still disrupted
strongly.

A qualitatively similar behaviour to the one discussed above for the
\texttt{TNTYST} EOS applies also for the stiffer
\texttt{BHB}$\Lambda\Phi$ EOS, although all of the dynamics is somewhat
milder and the disruptions weaker already at a mass criticality of
$M_{_{\infty}} / M^{^{1,0}}_{\rm th} = 1.05$. This points out that the
stellar compactness, and hence the EOS, also plays an important role in
the subtle balance that is introduced near the merger between tidal
forces (controlled by the mass ratio), centrifugal forces (controlled by
the primary spin), and the overall strength of the gravitational
potential (controlled by the total mass of the system). The most robust
dynamical features of this picture can therefore be summarised as
follows. First, in the equal-mass case the disruption is suffered by the
highly spinning primary, as spin effects dominate in this case. Second,
in the strongly mass asymmetric case, the disruption is instead suffered
by the secondary, as tidal effects dominate in this case. Finally, in all
cases, the tidal tails from the highly spinning primary carry large
amounts of angular momentum, expanding quicker, potentially shocking
partly into the tidal tail of the secondary.

\begin{figure*}
  \centering
  \centering
  \includegraphics[width=0.99\textwidth]{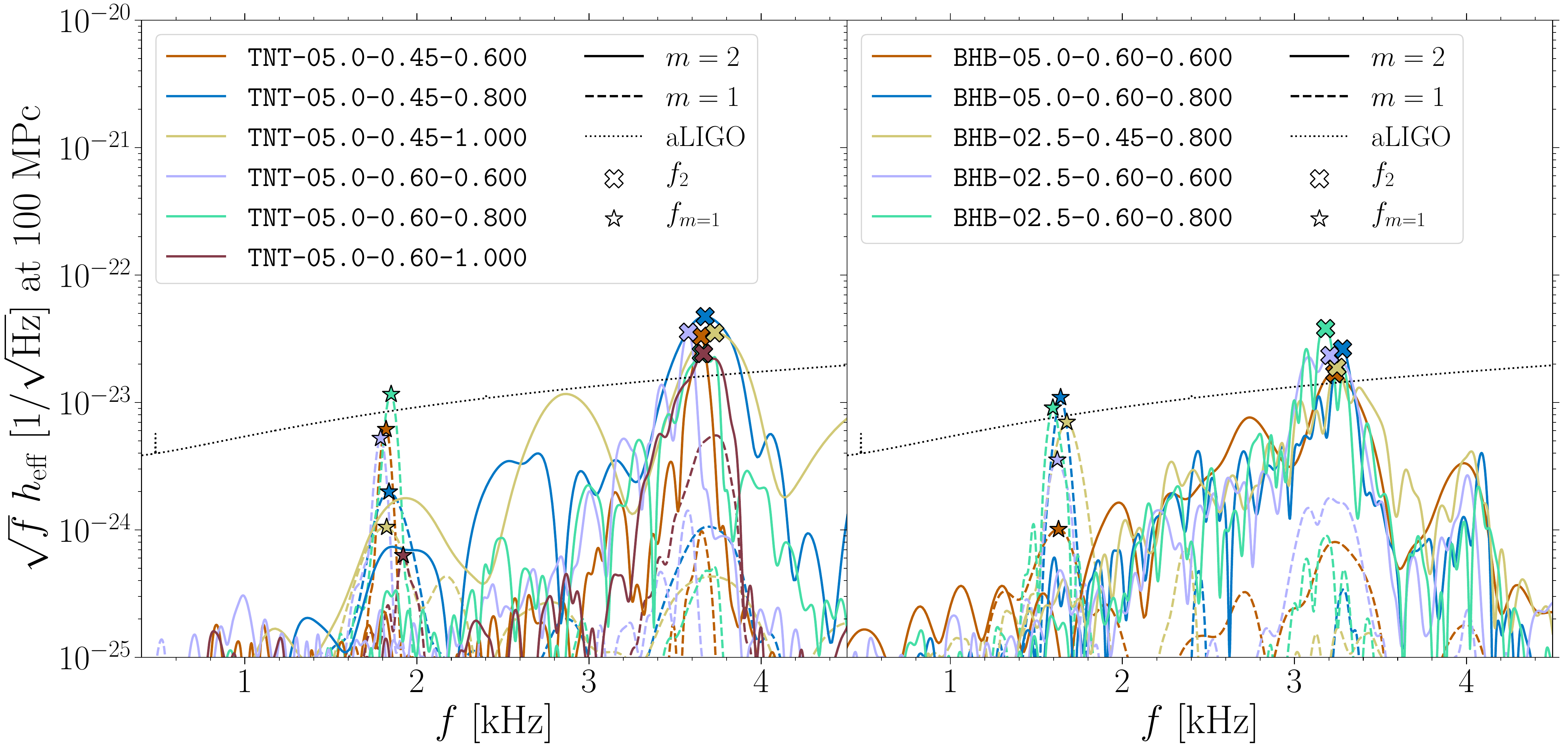}
  \caption{\textit{Left panel:} GW-strain spectral density for a number
    of binaries with either the \texttt{TNTYST} (\textit{left}) EOS. The
    spectra are computed over the time window between the merger and the
    collapse or till the end of the simulation; hence, they are different
    for different binaries. Marked with symbols are the positions of the
    dominant post-merger $f_{2}$ and of the one-arm instability $f_{m=1}$
    frequencies, which are produced by $\left(\ell=2,m=2 \right)$ and
    $\left(2{,}1 \right)$ deformations, respectively; shown with a dotted
    line is the sensitivity curve of Advanced LIGO. \textit{Right panel:}
    The same as in the left panel but for the \texttt{BHB}$\Lambda\Phi$
    EOS.}
  \label{fig:gw-spec}
\end{figure*}

Further into the post-merger phase, the merger remnants feature various
degrees of asymmetry in the matter distribution, especially in the
equatorial plane. Some representative examples are shown in
Fig.~\ref{fig:one-arm-tog} for two binaries with the \texttt{TNTYST} EOS
being equally above the threshold mass ($5\%)$ and having the same mass
ratio ($q=0.8$), but differing in the spin magnitude of the primary. Note
how the asymmetry of the matter distribution is quite significant and is
actually larger in the binary with the largest spin, \ie $\chi_1=0.6$. In
both cases, a strong one-arm instability (or $m=1$) develops in the
remnant \citep{Bernuzzi2013, Paschalidis2015, East2016, East2016a,
  Lehner2016a, Radice2016a, East2019}, and survives for significant
amounts of time, providing potentially measurable imprints in the
GW emission and disc winds \citep{Nedora2019}.

We next discuss the dynamics of the binaries examined in terms of their
GW emission. While the inspiral leads to different merger
times and to a different frequency evolution depending on the total mass,
mass ratio, spin magnitude of the primary, and the tidal deformability,
we focus on the post-merger remnant evolution. Of course, our
discussion will necessarily exclude those binaries that are not at least
short-lived (see Sec.~\ref{sec:results-stability} for some definitions).

To this scope, the spectrum of the post-merger GW emission for the
\texttt{TNTYST} EOS (left panel) and for the \texttt{BHB}$\Lambda\Phi$
EOS are shown in Fig.~\ref{fig:gw-spec}, displaying the frequencies of
the $\left(\ell=2{,}m=2\right)$-mode, together with the frequencies of the
$\left(2{,}1\right)$-mode. The spectrum of both modes is computed for the
effective strain over a window starting at merger until the system
undergoes a collapse or until the end of the simulation, in case of the
models with a sufficiently long post-merger HMNS phase \citep[see
  e.g.][for details]{Takami2014, Papenfort2018}. The windowing is done by
a Hann window to minimise spectral leakage from truncating the
signal. The resulting spectra are further smoothed by a spline
interpolation for presentation. The spectra show a clear contribution
from the $\left(2{,}1\right)$-mode with an expected frequency at $f_{m=1}
\approx 0.5\, f_{2}$ -- not to be confused with the $f_1$ mode in the
classification of \citet{Takami2014} -- where $f_{2}$ is the dominating
$\left(2{,}2\right)$-mode post-merger frequency \citep{Bauswein2011,
  Takami2014, Bernuzzi2015a, Palenzuela2015, Foucart2015, Lehner2016,
  Rezzolla2016}, confirming that the instability is triggered in all of
them to some extent.

Since the power in the $\left(2{,}1\right)$-mode reflects the asymmetry
in the mass of the system, its power depends sensitively on the mass
ratio, being almost a factor fifty smaller than the corresponding
$\left(2{,}2\right)$-mode in the case of equal-mass binaries (see left
panel). It should be noted, however, that the power at the $f_{m=1}$
frequency does not depend only on the strength of the $m=1$ deformation,
and hence on the mass ratio, but also on the survival time of the
remnant, which in turn depends on the spin of the binary.
For example, in the case of the \texttt{TNTYST} binaries with the largest
spin, \ie $\chi_{_{1}} = 0.6$, and which have a lifetime of at least
$34.8 \, \rm{ms}$ in all simulations, the strongest $m = 1$ contribution
is obtained for $q=0.8$ and not for $q=0.6$. As expected, even in this
specific high-spin case, the equal-mass binary has an $f_{m=1}$
contribution that is smaller than that of the unequal-mass $q=0.8$ case,
suggesting a non-monotonic response regarding the mass ratio. This
behaviour seems to suggest that there exist an optimal mass ratio for the
onset of the $m=1$ instability and that for very large mass asymmetries
and stiff EOSs, the mass of the secondary is too small to produce
significant $m=1$ deformations, thus leaving the $m=2$ bar-mode as the
dominant deformation.

This conjecture is supported by the results obtained with the
\texttt{BHB}$\Lambda\Phi$ binaries, even though the set of long-lived
models available is more limited. Also in these binaries, in fact, and
for the binaries with the largest primary spin $\chi_{_{1}} = 0.6$, the
instability is triggered more strongly at $q = 0.8$ compared to $q =
0.6$. This conclusion is more robust for binaries with a mass that is
$2.5 \%$ above the irrotational threshold mass, and for which both
models survive for at least $26.5 \, \rm{ms}$ after merger. The same
cannot be said for binaries with total masses at $5.0 \%$ above the
threshold mass, where the $q = 0.6$ binary unfortunately collapses
already $8.4 \, \rm{ms}$ after the merger.

\begin{figure*}
  \centering
  \centering
  \includegraphics[width=0.99\textwidth]{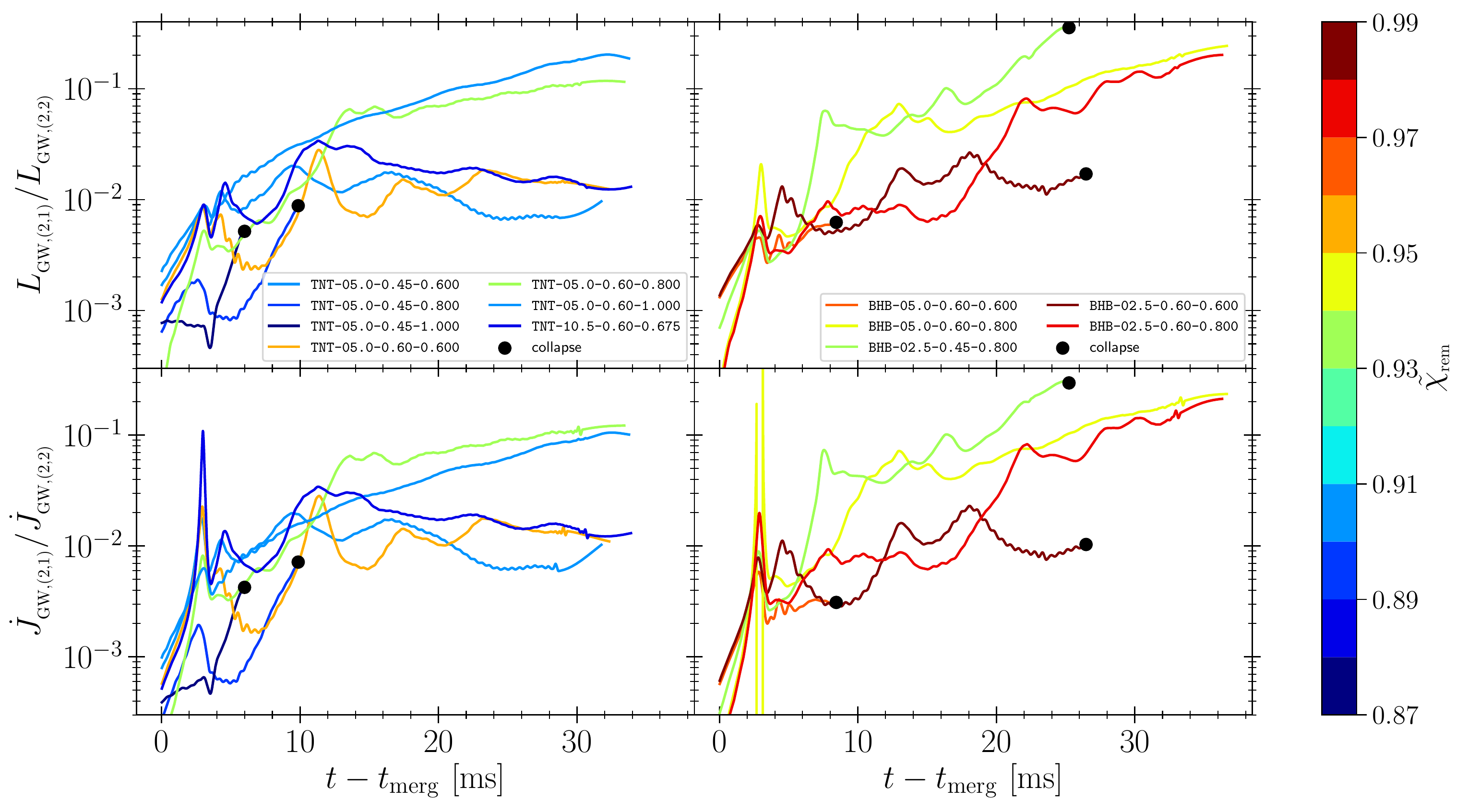}
  \caption{\textit{Left panel:} Ratio of the evolution of the
    instantaneous GW luminosities in the two modes, \ie $L_{{\rm GW},
      (2{,}1)}/L_{{\rm GW}, (2{,}2)}$ (top parts) and the corresponding
    losses of angular momentum, \ie $\dot{J}_{{\rm GW},
      (2{,}1)}/\dot{J}_{{\rm GW}, (2{,}2)}$ (bottom parts), for the
    \texttt{TNTYST} EOS (left panel). Black circles mark the time of
    collapse of the remnant. \textit{Left panel:} The same as in the left
    panel but for the \texttt{BHB}$\Lambda\Phi$ EOS.}
  \label{fig:GW_quantities}
\end{figure*}


We note that because the $m=1$ asymmetry introduced by the one-arm
instability tends to be more long-lived than the $m=2$ bar-mode asymmetry
triggered at merger by the large angular momentum of the remnant
\citep{Manca07}, the $m=1$ contribution grows for longer post-merger HMNS
lifetimes, increasing its contribution to the GW emission. Over time,
however, this growth levels off, so that the $m=1$ contribution remains
at values that are of the order of $\sim 1\%$ when compared to the $m=2$
emission over the timescales explored in our simulations. As an example,
simulating the $\texttt{BHB-02.5-0.60-0.800}$ binary at a lower
resolution of $295 \, \rm{m}$ revealed that the remnant is potentially
long-lived with a lifetime of at least $92 \, \rm{ms}$ when neglecting
any dissipative processes apart from GW radiation. In this case, the
ratio of the GW strain amplitudes in the $\left(2{,}1\right)$- and
$\left(2{,}2\right)$ -mode varies with time reaching $h_{21} / h_{22}
\approx 5$ at the end of the simulation, thus indicating that the $m=1$
asymmetry is long-lived \citep{Bernuzzi2013, Paschalidis2015,
  Radice2016a}.

Figure \ref{fig:GW_quantities} provides a synthetic overview of much of
what just discussed. More specifically, it shows the ratio of the
evolution of the instantaneous GW luminosities in the two modes, \ie
$L_{{\rm GW}, (2{,}1)}/L_{{\rm GW}, (2{,}2)}$ (top parts) and the
corresponding losses of angular momentum, \ie $\dot{J}_{{\rm GW},
  (2{,}1)}/\dot{J}_{{\rm GW}, (2{,}2)}$ (bottom parts), for the
\texttt{TNTYST} EOS (left panel) and for the \texttt{BHB}$\Lambda\Phi$
EOS (right panel).

The actual values are computed from the Weyl scalar $\Psi_4$ on a
spherical surface at a distance of $735 \, \rm km$ from the center of the
system \citep[see Eq. (117) and (119) in][]{Bishop2016} and the evolution
is deconvolved with a Savitzky-Golay filter \citep{Savitzky-Golay-1964} to
reduce the inevitable noise produced by the ratio of two rapidly varying
quantities.  Furthermore, black filled circles are used to mark the time
of collapse of the remnant. Note that the ratio of the luminosities grows
nonlinearly right after the merger and that,
although the evolution of these GW quantities is not correlated
uniquely with the rescaled dimensionless spin, $\tilde{\chi}_{\rm rem}$,
the ratios reach values of $\sim 0.01$ for binaries with small
$\tilde{\chi}_{\rm rem}$, and become $\sim 0.1$ for binaries with large
$\tilde{\chi}_{\rm rem}$. Furthermore, in some cases, and just before
collapse, $L_{{\rm GW}, (2{,}1)}/L_{{\rm GW}, (2{,}2)}$ can be as large
as $0.3$. Given the length over which the simulations have been carried
out, it remains unclear whether $L_{{\rm GW}, (2{,}1)}/L_{{\rm GW},
  (2{,}2)} \sim 1$ for those remnants with a lifetime spanning hundreds
of milliseconds. Future, long-term simulations of these binaries will
help settle this question.

To provide a more quantitative measure of the GW emissions for the
different binaries, Tab.~\ref{tab:ampl-ratios} reports the contributions
of the dominant post-merger frequency $f_2$ and the one-arm instability
frequency $f_{m=1}$ to the GW spectrum (see Fig.~\ref{fig:gw-spec}),
collectively within an $5 \, \rm{ms}$ time window after merger. In this
way, it is possible to note that while the strain amplitude in the
$f_{m=1}$ may dominate over long timescales, the effective power poured
into that mode at merger is nevertheless subdominant, being $\lesssim
8\%$ for all the cases surviving long enough to be considered. The data
in Tab. \ref{tab:ampl-ratios} also suggests that while the stiffness of
the EOS plays a major role in the dynamics of the $m=1$ instability, it
also shows that a highly spinning primary can change the response in the
newly formed HMNS contrary to expectation from irrotational binary
mergers. Especially the differences between $\chi_{_{1}} = 0.45$ and
$\chi_{_{1}} = 0.6$ for \texttt{TNTYST} suggest the development of this
instability depends sensitively on the combination of stiffness, spin and
mass asymmetry. Moreover, the ratio between the two contributions,
${h_{\mathrm{eff},f_{m=1}}}/{h_{\mathrm{eff},f_{2}}}$, changes over time,
as can be seen in Tab.~\ref{tab:ampl-ratios}, thus suggesting different
damping timescales for the $m=2$ and the $m=1$ modes.

\begin{table}
\begin{centering}
\addtolength{\tabcolsep}{-5pt}

\begin{tabular}{c|c|c|c|c|c|c}
  binary model
  & $h_{\mathrm{eff},f_2}$
  & $h_{\mathrm{eff},f_{m=1}}$
  & $\frac{h_{\mathrm{eff},f_{m=1}}}{h_{\mathrm{eff},f_{2}}}$
  & $\left.\frac{h_{\mathrm{eff},f_{m=1}}}{h_{\mathrm{eff},f_{2}}}\right|_{5 {\rm ms}}$
  & $E_{\rm GW,tot}$
  & $J_{\rm GW,tot}$

 \\ & & & & & $[M_{\odot}]$ & $[M^2_{\odot}]$\\

\hline 
\hline 
\scriptsize{\texttt{TNT-10.5-0.60-0.675}} & $ 0.500 $ & $ 0.032 $ & $ 0.064 $ & $ 0.173 $ & $ 0.05 $ & $ 0.97 $ \tabularnewline
\scriptsize{\texttt{TNT-10.5-0.60-0.838}} & $ 0.764 $ & $ 0.028 $ & $ 0.036 $ & $ 0.037 $ & $ 0.07 $ & $ 1.31 $ \tabularnewline
\hline                                                                       
\scriptsize{\texttt{TNT-05.0-0.45-0.600}} & $ 0.414 $ & $ 0.032 $ & $ 0.078 $ & $ 0.131 $ & $ 0.03 $ & $ 0.60 $ \tabularnewline
\scriptsize{\texttt{TNT-05.0-0.45-0.800}} & $ 0.682 $ & $ 0.023 $ & $ 0.034 $ & $ 0.030 $ & $ 0.08 $ & $ 1.54 $ \tabularnewline
\scriptsize{\texttt{TNT-05.0-0.45-1.000}} & $ 1.000 $ & $ 0.019 $ & $ 0.019 $ & $ 0.021 $ & $ 0.07 $ & $ 1.35 $ \tabularnewline
\hline                                                                       
\scriptsize{\texttt{TNT-05.0-0.60-0.600}} & $ 0.428 $ & $ 0.028 $ & $ 0.065 $ & $ 0.104 $ & $ 0.03 $ & $ 0.67 $ \tabularnewline
\scriptsize{\texttt{TNT-05.0-0.60-0.800}} & $ 0.594 $ & $ 0.033 $ & $ 0.056 $ & $ 0.341 $ & $ 0.05 $ & $ 1.09 $ \tabularnewline
\scriptsize{\texttt{TNT-05.0-0.60-1.000}} & $ 0.895 $ & $ 0.031 $ & $ 0.035 $ & $ 0.019 $ & $ 0.06 $ & $ 1.33 $ \tabularnewline
\hline                                                                       
\hline                                                                       
\scriptsize{\texttt{BHB-05.0-0.60-0.600}} & $ 0.333 $ & $ 0.019 $ & $ 0.057 $ & $ 0.042 $ & $ 0.02 $ & $ 0.52 $ \tabularnewline
\scriptsize{\texttt{BHB-05.0-0.60-0.800}} & $ 0.371 $ & $ 0.029 $ & $ 0.079 $ & $ 0.295 $ & $ 0.03 $ & $ 0.84 $ \tabularnewline
\hline                                                                       
\scriptsize{\texttt{BHB-02.5-0.45-0.800}} & $ 0.404 $ & $ 0.030 $ & $ 0.075 $ & $ 0.267 $ & $ 0.03 $ & $ 0.76 $ \tabularnewline
\hline                                                                       
\scriptsize{\texttt{BHB-02.5-0.60-0.600}} & $ 0.377 $ & $ 0.021 $ & $ 0.056 $ & $ 0.108 $ & $ 0.03 $ & $ 0.66 $ \tabularnewline
\scriptsize{\texttt{BHB-02.5-0.60-0.800}} & $ 0.428 $ & $ 0.027 $ & $ 0.063 $ & $ 0.169 $ & $ 0.04 $ & $ 0.96 $ \tabularnewline

\end{tabular}

\par\end{centering}
\caption{Contribution to the GW spectrum of the dominant post-merger
  frequency $f_2$ (second column), of the one-arm instability frequency
  $f_{m=1}$ (third column). Shown also are the ratio between these two
  quantities either over the whole lifetime (fourth column) or when
  integrated over a time window of $5 \, \rm{ms}$ after merger (fifth
  column). Since all the binaries reported lead to remnants that are
  either medium- or long-lived, the window is chosen so as to cover all
  models equally. The values are normalized by the highest power in this
  summary, \ie by the $f_2$-power of \texttt{TNT-05.0-0.45-1.000}.
  In addition the total post-merger radiated mass-energy $E_{\rm GW,tot}$
  (sixth column) and angular momentum $J_{\rm GW,tot}$ (seventh column)
  are listed, measured up to collapse respectively the end of the
  simulation.}

\label{tab:ampl-ratios}
\end{table}

\subsection{Asymmetry and lifetime of the remnant}
\label{sec:results-stability}

In the following we will discuss the lifetime of the merger remnant
before it collapses to a BH, distinguishing the results obtained for the
\texttt{TNTYST} EOS from those of the \texttt{BHB}$\Lambda\Phi$ results.

We start by recalling that the remnants of binary mergers with masses well
above the maximum mass allowed by uniform rotation \citep{Breu2016} are,
with due exceptions, in a metastable equilibrium as they will eventually
collapse to a system composed of a BH and an accreting torus
\citep{Baiotti2016}. The lifetime of the remnant depends therefore on a
number of properties of the binary, most notably, the total mass, the
mass ratio, the spin of the constituents, and, of course, the EOS. The
shortest lifetime is the one in which the collapse takes place within one
free-fall timescale \citep{Koeppel2019,Tootle2021} and can be estimated
in a number of different ways \citep[see also][]{Agathos2019,
  Bauswein2020}. We here follow the methodology first developed by
\citet{Koeppel2019} and further refined to account for spin and
unequal-mass systems by \citet{Tootle2021}, in which the normalized
minimum of the lapse $\hat{\alpha}(t) = \alpha_{\rm min}(t) / \max \left(
\alpha_{\rm min}(t) \right)$ is used to compute the lifetime $\tau_c$ and
compares it to the freefall timescale. In this approach, the lifetime is
then computed as the interval between the time of merger (defined as the
one at which $\hat{\alpha} = 0.9$) and the time of black-hole formation
(defined as when $\hat{\alpha}=0.1$).

Once the timescale $\tau_c$ is computed, we classify the post-merger
lifetime in three main classes, with the understanding that this
classification is in large part arbitrary and useful mostly to set some
general qualitative behaviours. In particular, we consider the merger
remnant to be: \textit{(i) short-lived} if the remnant collapses over a
timescale $\tau_c \lesssim 5 \, \text{ms}$\footnote{Given that the
free-fall timescale is below one millisecond for standard values of
neutron-star compactness, short-lived remnants include also those
undergoing prompt collapse.}. \textit{(ii) medium-lived} if the remnant
collapses over a timescale $5 \lesssim \tau_c \lesssim 10 \, \text{ms}$.
\textit{(iii) long-lived} if the remnant collapses over a timescale
$\tau_c \gtrsim 10 \, \text{ms}$.

\begin{figure*}
  \centering
  \centering
  \includegraphics[width=0.99\textwidth]{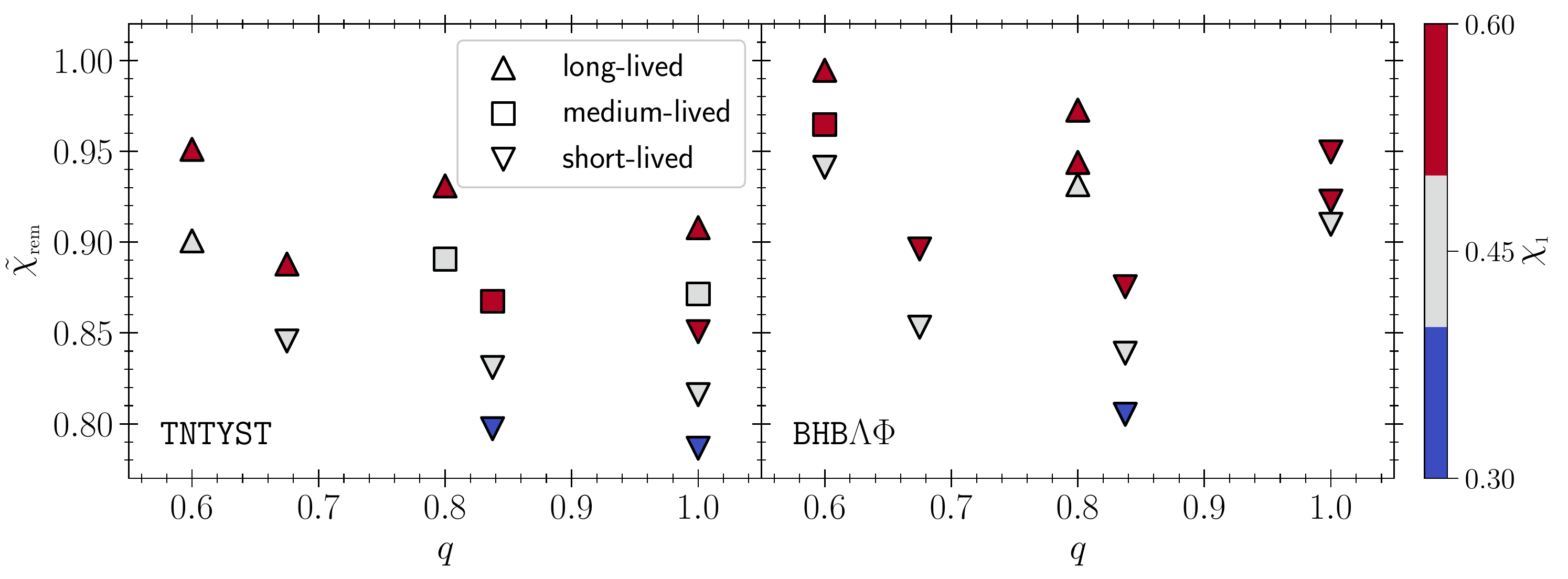}
  \caption{\textit{Left panel:} Rescaled dimensionless spin angular
    momentum of the remnant at merger [Eq. \eqref{eqn:al}] shown as a
    function of the mass ratio for the \texttt{TNTYST}
    binaries. Upward-pointing triangles refer to the long-lived remnants,
    squares to the medium-lived ones, while short-lived remnants are
    indicated with downward-pointing triangles (see
    Sec.~\ref{sec:results-stability} for a definition). Also indicated
    with a colourcode is the spin of the primary (see also
    Tab.~\ref{tab:initial}). \textit{Left panel:} The same as in the left
    panel but for the \texttt{BHB}$\Lambda\Phi$ EOS.}
  \label{fig:jm2-merger}
\end{figure*}

Concentrating on \texttt{TNTYST} configurations first, the long-lived
remnants that have not collapsed by the end of the simulations, survive
for at least $34.2 \, \rm{ms}$ after merger. This applies to models
$\texttt{TNT-10.5-0.60-0.675}, \texttt{TNT-05.0-0.45-0.600}$,
$\texttt{TNT-05.0-0.60-0.600}, \texttt{TNT-05.0-0.60-0.800}$, and
$\texttt{TNT-05.0-0.60-1.000}$, which, despite being $10.5 \%$ and $5.0
\%$ over $M^{^{1,0}}_{\rm th}$, do not show evidence of a collapse to a
BH over these timescales. Additionally, the three models
$\texttt{TNT-10.5-0.60-0.837}$, $\texttt{TNT-05.0-0.45-0.800}$ and
$\texttt{TNT-05.0-0.45-1.000}$, lead to medium-lived remnants, undergoing
a collapse after a timescale of $\tau_c \simeq 5.7 \, \text{ms},\ 9.9 \,
\text{ms}$, and $6.0 \, \text{ms}$ after merger, respectively. Finally,
model $\texttt{TNT-10.5-0.45-0.675}$ results in a short-lived post-merger
remnant ($\tau_c \simeq 2.4 \, \text{ms}$), indicating that this system is
very close to a prompt collapse in terms of total mass.

To quantify the effect of the spin angular momentum of the primary on the
lifetime of the remnant, we calculate the rescaled dimensionless angular
momentum of the remnant at merger, which we define as 
\begin{equation}
  \label{eqn:al}
  \tilde{\chi}_{_{\rm rem}} := \left(\frac{M^{^{1,0}}_{\rm
      th}}{M_{_{\infty}}} \right) \frac{J_{_{\rm ADM}} - J_{_{\rm
        GW}}}{\left(M_{_{\rm ADM}} - M_{_{\rm GW}}\right)^2} \,,
\end{equation}
where $J_{_{\rm ADM}}$ is the total ADM angular momentum and $M_{_{\rm
    ADM}}$ the total ADM mass of the initial data. Furthermore, since the
system is losing both angular momentum, $J_{_{\rm GW}}$, and
(gravitational) mass, $M_{_{\rm GW}}$, via the emission of GWs, these
need to be removed from the estimate of $\tilde{\chi}_{_{\rm rem}}$.
Both $J_{_{\rm GW}}$ and $M_{_{\rm GW}}$ are computed from the Weyl
scalar $\Psi_4$ on a spherical surface at a distance of $735 \, \rm km$
from the center of the system \citep[see \eg][]{Bishop2016} and are
integrated in time from the start of the simulations and up to the
merger. Because of this, $J_{_{\rm GW}}$ and $M_{_{\rm GW}}$ do
not contain the contributions radiated before the start of the
simulation, instead these are already correctly incoporated within
$J_{_{\rm ADM}}$ and $M_{_{\rm ADM}}$ by the underlying assumption of
quasi-circularity of the initial data. In addition, we employed a
rescaling by the mass criticality ${M^{^{1,0}}_{\rm
    th}}/{M_{_{\infty}}}$ to introduce a dependency of
super-criticality in the mass of the binary. While this choice is
arguably arbitrary, it has been introduced to reflect the fact that the
various models considered are above the irrotational threshold mass at
different degree, and this has an impact on the survival time of the
merger remnant quite independently of the magnitude of the angular
momentum at merger alone. Without this correction, it would be
difficult to distinguish the contribution to the remnant lifetime that
is introduced by the large spin of the system and by a comparatively
small mass; indeed, without this correction, Fig. \ref{fig:jm2-merger}
would not show a distinct systematic behaviour it shows now (see below
for a detailed discussion).

All in all, the total dimensionless angular momentum of the remnant given
by expression \eqref{eqn:al} provides an estimate of how much angular
momentum is in the remnant object at the time of the merger. Since
$J_{_{\rm GW}}$ and $M_{_{\rm GW}}$ can be estimated reasonably well with
post-Newtonian expressions, $\chi_{_{\rm rem}}$ is in principle a
quantity that can be estimated directly from the observations.

As already noted in Sec.~\ref{sec:results-post-merger}, the systems
experience longer inspirals with increasing spin of the primary due to
the spin-orbit coupling and the net increased angular momentum. This not
only leads to differing merger times for the same total mass
$M_{_{\infty}}$, but also to different $\chi_{_{\rm rem}}$ at merger
\citep[see also][]{Dietrich:2015b, Dietrich2017c}. A synthetic view of
the total dimensionless angular momentum at merger is presented for both
EOSs and as a function of the mass ratio in Fig.~\ref{fig:jm2-merger},
where the upward-pointing triangles refer to the long-lived remnants, the
squares to the medium-lived ones, while short-lived remnants are
indicated with downward-pointing triangles. Also indicated with a
colourcode is the spin of the primary. As expected, the systems with
primary stars having larger $\chi_{_{1}}$ show an increased $\chi_{_{\rm
    rem}}$ at merger.

In the case of the softer \texttt{TNTYST} EOS, the left panel of
Fig.~\ref{fig:jm2-merger} illustrates rather clearly that the spin in the
primary has the net effect of increasing the lifetime of the remnant, so
that all long-lived remnants are those having large values of
$\chi_1$. Interestingly, the spin of the remnant as defined by
Eq. \eqref{eqn:al} seems to cluster approximately into three different
regions: $\tilde{\chi}_{_{\rm rem}} \gtrsim 0.90$ for long-lived remnants,
$0.86 \lesssim \tilde{\chi}_{_{\rm rem}} \lesssim 0.90$ for remnants that
can be either long- or medium-lived , and $\tilde{\chi}_{_{\rm rem}}
\lesssim 0.86$ for remnants that are short-lived.
We also note that, in contrast to what has been found by concentrating on
the threshold mass of irrotational binaries with significant mass ratio
\citep{Bauswein2017b, Bauswein2020,Bauswein2020c,Bernuzzi2020} and in
line with the findings in \citet{Tootle2021} for $\left| \chi_{_{1}}
\right| \leq 0.3$, we find that with a significantly spinning primary,
the effective threshold mass grows with increasing mass asymmetry for
\texttt{TNTYST} (see the survival times in Tab.~\ref{tab:initial}).

What discussed so far continues to hold also for the stiffer
\texttt{BHB}$\Lambda\Phi$ EOS, although not without some important
differences. Also in this case, in fact, all long-lived remnants follow
from binaries having a rapidly spinning primary. Furthermore, the spin of
the remnant at merger is again larger in the presence of large mass
asymmetries. However, there are notable differences comparing the right
to the left panel of Fig.~\ref{fig:jm2-merger}. First,
$\tilde{\chi}_{_{\rm rem}}$ exhibits overall a larger spread, second,
there is a clear exception from the trend identified above with the
\texttt{TNTYST} EOS, namely, that for equal-mass binaries a very rapidly
spinning primary does not prevent the merger remnant to be
short-lived. The fact that this behaviour is found for the
\texttt{BHB}$\Lambda\Phi$ EOS only indicates that the impact of the spin
is less important in equal-mass mergers with stiff EOSs. On the other
hand, what is even more relevant for this EOS is that it is much harder
to find binaries with medium-lived remnants. Indeed, the simulated
systems exhibit mostly remnants that are either short-lived (nine out of
14 binaries) or long-lived (four out of 14 binaries). Only in one
instance, \ie \texttt{BHB-05.0-0.60-0.600}, does the remnant survive for
$8.4\, \text{ms}$ before collapsing. Not surprisingly, the few long-lived
remnants, \ie the models $\texttt{BHB.050.0.60-0.800}$,
$\texttt{BHB.025.0.45-0.800}$, $\texttt{BHB.025.0.60-0.600}$ and
$\texttt{BHB.025.0.60-0.800}$ (the corresponding HMNSs survive at least
for $25.2 \, \text{ms}$ for these models), also correspond to those
binaries that have masses only up to $5\,\%$ above the threshold mass for
irrotational binaries. This result underlines how the lifetime of the
remnant can be increased if the system has large mass asymmetries, a
primary with a large spin, and is not excessively massive. Indeed, for
the binary $\texttt{BHB.025.0.60-0.800}$, the remnant shows long-term
lifetime of at least $92.0 \, \text{ms}$ in the lower-resolution
simulation (see discussion in Sec.~\ref{sec:results-post-merger}).

As anticipated in Sec.~\ref{sec:results-post-merger}, the lifetime of the
merger remnant has an impact on the GW spectrum and on the relative
importance of the $m=1$ and $m=2$ deformations. At the same time, reading
off such importance from the amplitude of the $f_{m=1}$ and $f_2$ peaks
in Fig.~\ref{fig:gw-spec} can be misleading, because they reflect
different lifetimes of the merger remnant. This can be accommodated by
considering the emitted power within a precise window in time, \eg $5 \,
\rm{ms}$ as reported in Tab.~\ref{tab:ampl-ratios}, as in this case all
models are evaluated equally. Doing so highlights that for the
\texttt{TNTYST} EOS models, a smaller mass ratio (and therefore a larger
mass asymmetry) leads to a comparatively larger $m=1$ emission and to
longer lifetimes\footnote{We note that, strictly
speaking, the equal-mass binaries with the \texttt{TNTYST} EOS are also
those that lead to the largest amounts of radiated GW energy (see
Tab.~\ref{tab:ampl-ratios}). These binaries, however, are short lived and
the $m=1$ does not have sufficient time to provide a large contribution
as it is the case for binaries with larger mass asymmetry.}.
This indicates that --
for a given mass and spin -- a small mass ratio leads to a weakening of
the $m=2$ deformation, which, in turn, produces a reduced loss of GWs and
angular momentum, and thus a longer lifetime. At the same time -- for a
given mass and mass ratio -- a larger spin unambiguously increases the
remnant lifetime (see the values of $\tau_c$ in Tab.~\ref{tab:initial}).

\begin{figure*}
  \centering
  \centering
  \includegraphics[width=0.9\textwidth]{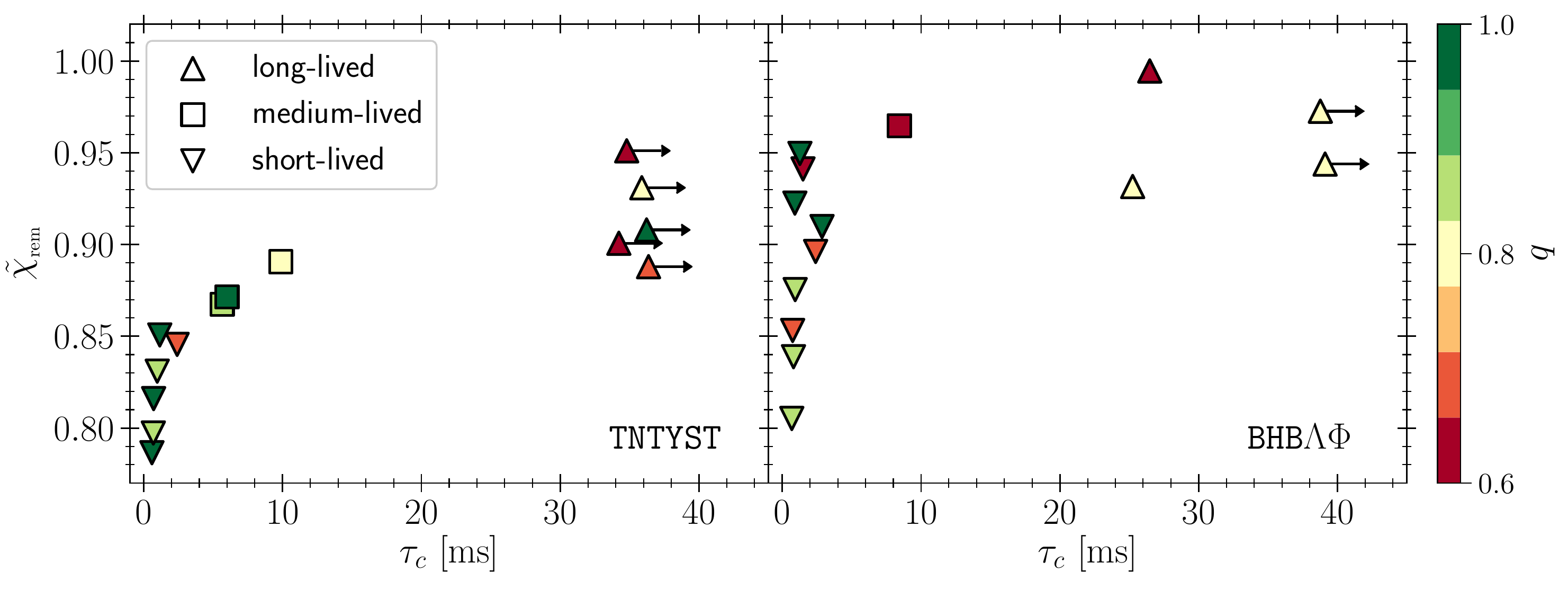}
  \caption{\textit{Left panel:} Rescaled dimensionless spin angular
    momentum of the remnant at merger [Eq. \eqref{eqn:al}] shown as a
    function of the remnant lifetime $\tau_c$ for the \texttt{TNTYST}
    binaries. Upward-pointing triangles refer to the long-lived remnants,
    squares to the medium-lived ones, while short-lived remnants are
    indicated with downward-pointing triangles (see
    Sec.~\ref{sec:results-stability} for a definition). Also indicated
    with a colourcode is the mass ratio $q$ of the binary. \textit{Left
      panel:} The same as in the left panel but for the
    \texttt{BHB}$\Lambda\Phi$ EOS.}
  \label{fig:jm2-merger_tau}
\end{figure*}

The role played by the mass asymmetry in the merger remnant and the
impact it has on its lifetime can be appreciated from
Fig.~\ref{fig:one-arm-tog}, when comparing the rest-mass density of the
binaries $\texttt{TNT-05.0-0.45-0.800}$ (top panels) and
$\texttt{TNT-05.0-0.60-0.800}$ (bottom panels), which differ only in the
spin of the primary. These two binaries show a large difference in both
the HMNS lifetime (which is at four times larger for $\chi_1=0.6$ than
for $\chi_1=0.45$, with the former not collapsing until the end of the
simulation) and the ratio of the $m=1$ and $m=2$ GW emissions (which is
of factor 10 within the first $5\,{\rm ms}$, that is, almost twice as
large for $\chi_1=0.6$ than for $\chi_1=0.45$). In both cases, the
remnant exhibits an asymmetry in the density distribution, which however
are dissipated more rapidly in the less-spinning binary $\chi_{_{1}} =
0.45$, whose core assumes an almost axisymmetric distribution after about
$9\, {\rm ms}$. On the other hand, the mass asymmetry is more long-lived
for the highly spinning binary with $\chi_{_{1}} = 0.6$ and is indeed
present until the end of the simulation at $t-t_{\rm mer} = 35.9 \,
\rm{ms}$. Clearly, the asymmetry in this case will lead to a persistent
GW emission in the $m=1$ mode.

A similar behaviour is exhibited also by the binaries of the
\texttt{BHB}$\Lambda\Phi$ EOS, with the most relevant difference being
that the lifetime of the remnant in this case is maximised not for the
binaries with the smallest mass ratio, but rather with an intermediate
one. Once again, this is related to the ability of triggering a
long-lasting asymmetry in the remnant, able to produce GW at a reduced
rate and thus, in conjunction with larger $\tilde{\chi}_{\rm rem}$,
increase its lifetime.

In summary, the results presented in this section suggest that the
properties of the merger remnant, most notably its degree of asymmetry in
the rest-mass distribution and its lifetime, depend on a subtle balance
between the mass of the binary, the mass ratio, and the spin of the
primary. In this complex combination, long-lived remnants tend to be
systematically associated to large dimensionless spins at merger. However,
while for binaries with softer EOSs this happens at the smallest mass
ratios, namely, at $q=0.6$ for the \texttt{TNT} EOS, for binaries with
stiffer EOSs this happens at an intermediate mass ratio, namely, at
$q=0.8$ for the \texttt{BHB}$\Lambda\Phi$ EOS.

All of this is nicely summarised in Fig. \ref{fig:jm2-merger_tau}, which
reports the rescaled dimensionless spin angular momentum at merger
$\tilde{\chi}_{\rm{rem}}$ for the \texttt{TNTYST} (\textit{left}) and the
\texttt{BHB}$\Lambda\Phi$ (\textit{right}) binaries shown as a function
of the remnant lifetime $\tau_c$. The symbol notation is the same as in
previous figures (see Table \ref{tab:initial}) and distinguishes between
long-, medium- and short-lived remnants; also indicated with a colourcode
is the mass ratio $q$ of the binary. Note how both panels indicate that
the lifetime increases with the dimensionless spin and that this
dependence can be considerably nonlinear especially for remnants that are
short- or medium-lived. Furthermore the long-lived remnants are also
those with the largest dimensionless spin at merger, thus pointing out
that the ability of storing angular momentum in the remnant represents
the most efficient way to extend the lifetime of the merger
product. Finally, note that the longest lived remnants are normally
those with the smallest mass ratio in the case of the soft EOS
\texttt{TNTYST}, \ie $q=0.6$. However, this is not the case for the
stiffer \texttt{BHB}$\Lambda\Phi$ EOS, where the longest-lived remnants
are actually those with intermediate mass ratio, \ie $q=0.8$. This is an
interesting demonstration that, as mentioned above, the lifetime of the
merger remnant is a complex balance between mass, mass ratio, spin, and
EOS properties. While this is true also for the threshold mass to prompt
collapse $M_{\rm th}$, where the work of \citet{Tootle2021} has shown
that it is even possible to describe a quasi-universal behaviour of
$M_{\rm th}$ as a function of $q$ and $\chi_1$, the properties of the
remnant lifetime appear more complex and hence more difficult to
describe in terms of a simple quasi-universal behaviour. This is not
particularly surprising given the complex nonlinear dynamics that follows
the merger and the development of instabilities of various type, some of
which, (\eg those associated with the presence of strong magnetic fields)
are ignored here.

\begin{figure*}
  \centering \centering
  \includegraphics[width=0.95\columnwidth]{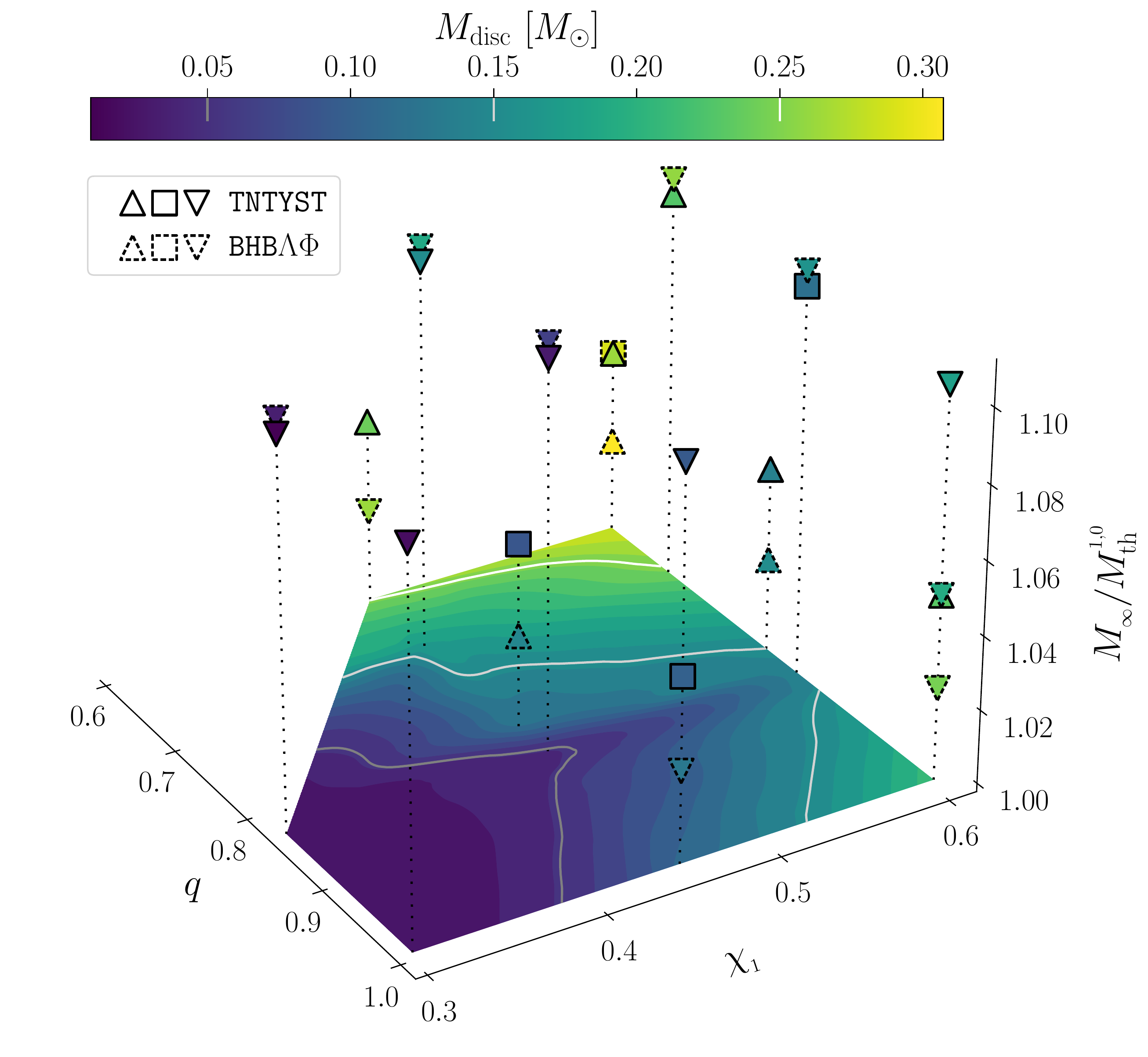}
  \includegraphics[width=0.95\columnwidth]{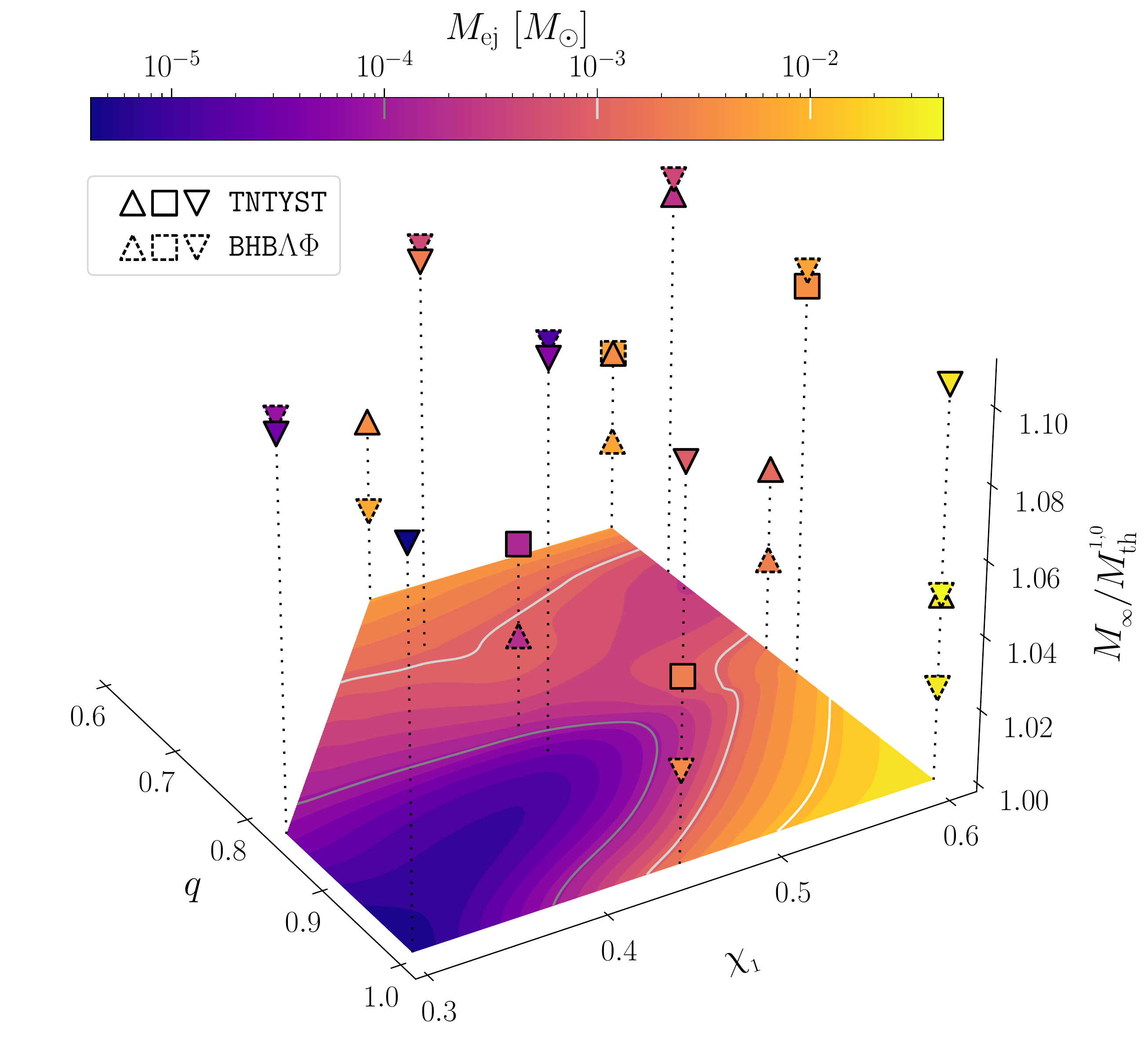}
  \caption{\textit{Left panel:} Remnant disc masses, $M_{\rm disc}$, for
    both EOSs shown with a colourcode as function of the dimensionless
    spin of the primary $\chi_{_{1}}$, of the mass ratio $q$, and of the
    the mass criticality $M_{_{\infty}} / M^{^{1,0}}_{\rm th}$ (the
    latter represents the vertical axis). The different symbols follow
    the convention presented in Tab.~\ref{tab:initial} and refer to the
    different lifetime of the remnants. Shown on the horizontal plane is
    a cubic spline interpolation of the data, where an average is
    performed for all models at fixed mass ratio and primary spin. The
    projection also reports isocontours of $M_{\rm disc}$ at $0.05, 0.15,
    0.25\,M_{\odot}$. \textit{Right panel:} The same as the left but
    reporting the dynamically ejected mass $M_{\rm ej}$. The projection
    on the horizontal plane reports isocontours of $M_{\rm ej}$ at
    $10^{-4},10^{-3},10^{-2}\,M_{\odot}$.}
  \label{fig:disc-masses}
\end{figure*}

\subsection{Dynamical ejecta and remnant disc mass}
\label{sec:results-matter}

We next discuss how to use the simulations to estimate the remnant disc
-- both when the merger product collapses to a BH and when it is
long-lived -- and the dynamically ejected mass for of all binaries
considered. Both of these quantities are essentials to put constraints
on the observable characteristics of these systems and hence on their
multi-messenger appearance. Numerical values for the disc masses can be
found in Tab.~\ref{tab:initial}, while the dynamical ejecta together with
their average properties are collected in Tab.~\ref{tab:ejecta-kn}.

We have estimated the disc masses via two different methods,
depending on whether the remnant undergoes a collapse to a BH or stays
metastable within the given simulation times. In the former case, the
disc mass is calculated by integrating over the conserved rest-mass
density except for the volume within the apparent horizon; in the latter
case, instead we select a cut-off rest-mass density of $10^{13} \, \rm{g
  / cm^3}$ and thus exclude from the integral the matter contribution
within the HMNS \citep[see \eg][]{Hanauske2016,Radice2018a}. We note that
because matter is still undergoing accretion onto the BH or is settling
into a quasi-steady state around the HMNS after the merger, we compute
the disc masses either $5 \, \rm{ms}$ after collapse in the first
scenario, or $15 \, \rm{ms}$ after merger in the second one. Of course
this choice is somewhat arbitrary and brings in a certain error, as the
disc mass estimates reach a genuine stationary value only on much longer
time-scales, where disc-ejection mechanisms will start to dominate
\citep[see][for a review]{Metzger2017,Gill2019}. We estimate this error
in the disc mass related to this choice to be $\approx 10-20 \%$.

The left panel of Fig.~\ref{fig:disc-masses} shows with a colourcode the
disc masses $M_{_{\rm disc}}$ as a function of the primary spin
$\chi_{_{1}}$, of the mass ratio $q$, and of the mass criticality
$M_{_{\infty}} / M^{^{1,0}}_{\rm th}$, shown as the vertical axis. The
symbol notation is the same as in Tab.~\ref{tab:initial} and is used to
distinguish short-, medium- and long-lived remnants, with solid symbols
referring to the \texttt{TNTYST} EOS, while dashed ones are used for the
\texttt{BHB}$\Lambda\Phi$ EOS.

A number of interesting behaviours appear quite clearly from the left
panel of Fig. \ref{fig:disc-masses} and apply equally to the
\texttt{TNTYST} and \texttt{BHB}$\Lambda\Phi$ EOSs. First, the smallest
disc masses are attained for binaries having low spin in
the primary; clearly, even smaller disc masses would be expected for
binaries with $\chi_1=0.0$ (not shown in
Fig. \ref{fig:disc-masses}). Second, and by contrast, the largest disc
masses are reached for binaries having small mass ratios and high spin in
the primary; under these conditions, discs with masses well above
$0.1\,M_{\odot}$ are possible for both EOSs and would lead to a
significant contribution to the matter ejected secularly since some of
these models lead to a long-lived remnant (see discussion in
sec. \ref{sec:results-kn}). Third, the disc masses can be considerably
different between the high- and low-spin systems, varying by almost three
orders of magnitude for the \texttt{TNTYST} EOS and the binary with mass
criticality $M_{_{\infty}} / M^{^{1,0}}_{\rm th} = 1.105$. Fourth, for
low-spin binaries, \ie $\chi_1=0.3$, the variation in disc mass with mass
ratio is very mild and $\lesssim30\%$; this is not the case for rapidly
spinning primaries, \ie $\chi_1=0.45$, where the variation in disc mass
with $q$ can be $\gtrsim50\%$. Finally, and interestingly, a local
minimum appears for $q=0.8$ and highly spinning binaries, \ie
$\chi_1=0.6$, once again pointing out that the combined impact of mass
ratio and spin is not trivial and can lead to non-monotonic behaviours of
the disc mass in highly spinning binaries. An analytic modelling of the
disc masses as a function of the mass ratios and spins of the binaries
will be discussed in a forthcoming work.


Following in a similar fashion, the right panel of
Fig.~\ref{fig:disc-masses} provides a synthetic summary of the mass that
is ejected dynamically using the same convention employed in
Fig.~\ref{fig:disc-masses}; more precise numerical values of the ejected
mass can also be found in Tab.~\ref{tab:ejecta-kn} alongside with the
average electron fraction $\langle{Y}_e\rangle$, velocity at infinity
$\langle v \rangle_{\rm ej}$, and entropy per baryon $\langle s
\rangle$. The table also reports the total ejected mass $M_{\rm tot}$,
which we define as the sum of the dynamically ejected mass and of a
fraction of the rest-mass of the remnant disc, \ie
\begin{equation}
\label{eq:Mtot}
  M_{\rm tot} := M_{\rm ej}+\frac{1}{2}M_{\rm disc}\,,
\end{equation}
where the disc mass is expected to be ejected secularly via a number of
physical processes such as neutrino or magnetically driven winds
\citep[see][for a summary of these secular winds]{Gill2019}.

Also in the case of the dynamically ejected matter, a number of
interesting behaviours can be deduced from the right panel of
Fig. \ref{fig:disc-masses} and apply to both EOSs. First, the smallest values
of the ejecta are attained for binaries having equal masses and low spin
in the primary; clearly, even smaller ejected masses would be expected
for binaries with $q=1, \chi_1=0.0$ (not shown in
Fig. \ref{fig:disc-masses}). Second, considerably larger ejecta are reached
for binaries having small mass ratios and high spin in the primary; under
these conditions, the ejected mass can be of the order of
$10^{-3}\,M_{\odot}$ for both EOSs (see Tab.~\ref{tab:ejecta-kn}). Third,
and differing from what seen for the disc mass, the largest amounts of
ejected matter are attained for equal-mass, rapidly spinning binaries;
more specifically for $q=1$ and $\chi_1=0.6$, the dynamically ejected
mass can be of the order of $10^{-2}\,M_{\odot}$ for both EOSs, yielding
a total ejected mass in excess of $0.1\,M_{\odot}$ (see
Tab.~\ref{tab:ejecta-kn}). Finally, we report again evidence for a
non-monotonic behaviour of the ejected mass when moving across different
mass ratios and for binaries with fixed spin.

When examining the additional average properties of the ejecta presented
in Tab.~\ref{tab:ejecta-kn}, \eg the average electron fraction
$\langle{Y}_e\rangle$, the average velocity at infinity $\langle v
\rangle_{\rm ej}$, the average entropy $\langle s \rangle$ together with
the derived quantities, it is possible to appreciate that they lie within
the typical range for rather cold, very neutron-rich and fast ejecta,
typically found in quasi-circular BNS mergers \citep[see
  \eg][]{Radice2016,Bovard2017}. The most relevant difference can be
found in those configurations showing a comparatively larger entropy,
velocity or composition. In these cases, the tidal tails and remnant
matter from the tidal disruption, undergo a multitude of
interactions. Most notably, the matter ejected from the spinning primary
expands faster thanks to the inherent large angular momentum; in turn,
this implies that in some cases the expanded matter shocks with the
elongated matter tails produced by the tidal disruption, thus leaving
potential imprints on the properties of the ejecta.

\begin{table*}
\begin{centering}
\begin{tabular}{c|c|c|c|c|c|c||c|c|c}
binary model & $M_{_{\rm ej}}$ & $\langle{Y}_e\rangle$ & $\langle v \rangle_{\rm ej}$    & $\langle s \rangle$                               & $M_{_{\rm tot}}$ & $\langle v \rangle$            & $t_{p}$              & $L_{p}$      & $T_{p}$ \\
& $\left[10^{-2} \, M_\odot\right]$  &             & $\left[c\right]$  & $\left[k_{\rm B} / \rm{baryon} \right]$ & $\left[10^{-1} \, M_\odot\right]$ & $\left[\rm c\right]$ & $\left[\rm d\right]$ & $\left[10^{40} \, \rm erg/s\right]$ & $\left[\rm K\right]$ \\

\hline 
\hline 
\texttt{TNT-10.5-0.30-0.837} & $ 0.0031 $ & $ 0.14 $ & $ 0.20 $ & $ 18.9 $ & $ 0.047 $ & $ 0.101 $ & $ 0.78 $ & $ 2.03 $ & $ 3880 $ \tabularnewline
\texttt{TNT-10.5-0.30-1.000} & $ 0.0004 $ & $ 0.10 $ & $ 0.11 $ & $ 43.1 $ & $ 0.097 $ & $ 0.100 $ & $ 0.99 $ & $ 2.77 $ & $ 3725 $ \tabularnewline
\texttt{TNT-10.5-0.45-0.675} & $ 0.2131 $ & $ 0.08 $ & $ 0.21 $ & $ 8.1 $ & $ 0.783 $ & $ 0.103 $ & $ 1.89 $ & $ 6.89 $ & $ 3323 $ \tabularnewline
\texttt{TNT-10.5-0.45-0.837} & $ 0.0051 $ & $ 0.13 $ & $ 0.19 $ & $ 25.3 $ & $ 0.149 $ & $ 0.100 $ & $ 1.13 $ & $ 3.33 $ & $ 3639 $ \tabularnewline
\texttt{TNT-10.5-0.45-1.000} & $ 0.0843 $ & $ 0.04 $ & $ 0.13 $ & $ 5.8 $ & $ 0.469 $ & $ 0.101 $ & $ 1.63 $ & $ 5.43 $ & $ 3417 $ \tabularnewline
\texttt{TNT-10.5-0.60-0.675} & $ 0.0233 $ & $ 0.06 $ & $ 0.11 $ & $ 13.6 $ & $ (1.142) $ & $ (0.100) $ & $ (2.17) $ & $ (7.90) $ & $ (3254) $ \tabularnewline
\texttt{TNT-10.5-0.60-0.837} & $ 0.3597 $ & $ 0.04 $ & $ 0.14 $ & $ 4.9 $ & $ 0.627 $ & $ 0.102 $ & $ 1.77 $ & $ 6.23 $ & $ 3363 $ \tabularnewline
\texttt{TNT-10.5-0.60-1.000} & $ 2.6884 $ & $ 0.04 $ & $ 0.17 $ & $ 1.5 $ & $ 1.153 $ & $ 0.116 $ & $ 2.00 $ & $ 8.87 $ & $ 3257 $ \tabularnewline

\hline 
\texttt{TNT-05.0-0.45-0.600} & $ 0.3517 $ & $ 0.05 $ & $ 0.14 $ & $ 5.1 $ & $ (1.239) $ & $ (0.101) $ & $ (2.21) $ & $ (8.25) $ & $ (3239) $ \tabularnewline
\texttt{TNT-05.0-0.45-0.800} & $ 0.0156 $ & $ 0.13 $ & $ 0.16 $ & $ 21.2 $ & $ 0.444 $ & $ 0.100 $ & $ 1.60 $ & $ 5.29 $ & $ 3427 $ \tabularnewline
\texttt{TNT-05.0-0.45-1.000} & $ 0.2335 $ & $ 0.07 $ & $ 0.16 $ & $ 7.5 $ & $ 0.525 $ & $ 0.103 $ & $ 1.67 $ & $ 5.80 $ & $ 3397 $ \tabularnewline
\texttt{TNT-05.0-0.60-0.600} & $ 0.3365 $ & $ 0.05 $ & $ 0.14 $ & $ 5.1 $ & $ (1.347) $ & $ (0.101) $ & $ (2.27) $ & $ (8.54) $ & $ (3225) $ \tabularnewline
\texttt{TNT-05.0-0.60-0.800} & $ 0.1234 $ & $ 0.04 $ & $ 0.12 $ & $ 6.0 $ & $ (0.701) $ & $ (0.100) $ & $ (1.85) $ & $ (6.44) $ & $ (3342) $ \tabularnewline
\texttt{TNT-05.0-0.60-1.000} & $ 2.8262 $ & $ 0.04 $ & $ 0.16 $ & $ 1.6 $ & $ (1.438) $ & $ (0.112) $ & $ (2.18) $ & $ (9.53) $ & $ (3217) $ \tabularnewline
\hline 

\hline 
\texttt{BHB-10.9-0.30-0.837} & $ 0.0079 $ & $ 0.12 $ & $ 0.20 $ & $ 14.6 $ & $ 0.174 $ & $ 0.100 $ & $ 1.19 $ & $ 3.55 $ & $ 3609 $ \tabularnewline
\texttt{BHB-10.9-0.45-0.675} & $ 0.0448 $ & $ 0.04 $ & $ 0.11 $ & $ 6.3 $ & $ 0.942 $ & $ 0.100 $ & $ 2.04 $ & $ 7.28 $ & $ 3288 $ \tabularnewline
\texttt{BHB-10.9-0.45-0.837} & $ 0.0013 $ & $ 0.10 $ & $ 0.12 $ & $ 34.5 $ & $ 0.327 $ & $ 0.100 $ & $ 1.46 $ & $ 4.64 $ & $ 3485 $ \tabularnewline
\texttt{BHB-10.9-0.60-0.675} & $ 0.0430 $ & $ 0.05 $ & $ 0.12 $ & $ 9.4 $ & $ 1.300 $ & $ 0.100 $ & $ 2.26 $ & $ 8.35 $ & $ 3231 $ \tabularnewline
\texttt{BHB-10.9-0.60-0.837} & $ 0.5922 $ & $ 0.03 $ & $ 0.14 $ & $ 2.4 $ & $ 0.870 $ & $ 0.102 $ & $ 1.96 $ & $ 7.17 $ & $ 3304 $ \tabularnewline

\hline 
\texttt{BHB-05.0-0.60-0.600} & $ 0.6126 $ & $ 0.04 $ & $ 0.12 $ & $ 3.4 $ & $ 1.509 $ & $ 0.101 $ & $ 2.36 $ & $ 8.96 $ & $ 3204 $ \tabularnewline
\texttt{BHB-05.0-0.60-0.800} & $ 0.1959 $ & $ 0.03 $ & $ 0.12 $ & $ 3.9 $ & $ (0.812) $ & $ (0.100) $ & $ (1.94) $ & $ (6.85) $ & $ (3315) $ \tabularnewline
\texttt{BHB-05.0-0.60-1.000} & $ 4.2371 $ & $ 0.05 $ & $ 0.17 $ & $ 0.9 $ & $ 1.381 $ & $ 0.120 $ & $ 2.06 $ & $ 9.89 $ & $ 3226 $ \tabularnewline

\hline 
\texttt{BHB-02.5-0.45-0.600} & $ 0.7048 $ & $ 0.04 $ & $ 0.11 $ & $ 2.6 $ & $ 1.382 $ & $ 0.101 $ & $ 2.30 $ & $ 8.61 $ & $ 3220 $ \tabularnewline
\texttt{BHB-02.5-0.45-0.800} & $ 0.0189 $ & $ 0.14 $ & $ 0.13 $ & $ 24.0 $ & $ 0.732 $ & $ 0.100 $ & $ 1.88 $ & $ 6.54 $ & $ 3334 $ \tabularnewline
\texttt{BHB-02.5-0.45-1.000} & $ 0.3071 $ & $ 0.04 $ & $ 0.13 $ & $ 3.6 $ & $ 0.674 $ & $ 0.101 $ & $ 1.82 $ & $ 6.37 $ & $ 3350 $ \tabularnewline
\texttt{BHB-02.5-0.60-0.600} & $ 0.6491 $ & $ 0.04 $ & $ 0.12 $ & $ 3.2 $ & $ 1.601 $ & $ 0.101 $ & $ 2.40 $ & $ 9.19 $ & $ 3194 $ \tabularnewline
\texttt{BHB-02.5-0.60-0.800} & $ 0.2362 $ & $ 0.03 $ & $ 0.12 $ & $ 3.7 $ & $ (0.792) $ & $ (0.101) $ & $ (1.92) $ & $ (6.79) $ & $ (3320) $ \tabularnewline
\texttt{BHB-02.5-0.60-1.000} & $ 3.0570 $ & $ 0.04 $ & $ 0.17 $ & $ 1.0 $ & $ 1.537 $ & $ 0.113 $ & $ 2.22 $ & $ 9.86 $ & $ 3205 $ \tabularnewline

\end{tabular}
\par\end{centering}
\caption{Average properties of the dynamical ejecta and of the
  kilonova-emission characteristics for all of the binaries
  evolved. Listed are the the total dynamically ejected mass $M_{_{\rm
      ej}}$, the average electron fraction $\langle{Y}_e\rangle$, the
  average velocity at infinity $\langle v \rangle_{\rm ej}$, the average
  entropy $\langle s \rangle$, the total ejecta $M_{\rm tot}$, the
  mass-weighted ejecta velocity $\langle v \rangle$, the peak epoch of
  the kilonova emission $t_p$, the peak bolometric luminosity $L_p$, and
  the effective temperature at the peak $T_p$. The results for the
  long-lived remnants are given between round brackets due to the unknown
  remnant lifetimes.}

\label{tab:ejecta-kn}
\end{table*}

\subsection{Implications on the kilonova emission}
\label{sec:results-kn}

The results discussed in the previous section on the disc mass and
ejected matter have a direct impact on the electromagnetic counterpart to
the BNS merger and, in particular, on the kilonova emission. To discuss
such implications, it is useful to distinguish the case of short-lived
remnants from that of the long-lived ones.

In the first case, namely, for binaries leading to a short-lived remnant
but retaining a large amount of dynamical ejecta and a massive remnant
disc, the kilonova emission is expected to contain merely a red
component. Simulations concentrating on matter ejection from the disc
around a remnant BH have shown that up to $40 \%$ of the disc mass can
be ejected at average velocities of $\langle v \rangle \approx 0.1 \, c$
and electron fraction of $\langle{Y}_e\rangle \approx 0.2$
\citep{Siegel2017, Siegel2018, Fernandez2018, Christie2019b}. However, it
is not clear yet if these results with initial disc masses of $\approx
0.03 \, \Msol$ can be scaled to massive discs as those obtained in many
of our configurations. The low electron fraction and moderate velocities
of the dynamical ejecta found in our simulations for these configurations
argue for a pure red kilonova. Indeed, we find that the dynamically ejected
part of the outflow averages around $\langle v \rangle \approx 0.14$ and
$\langle{Y}_e\rangle \approx 0.06$, so that we assume here that both
components evolve as a single red component. Furthermore, an electron
fraction $Y_e \approx 0.2$ from the secular disc ejecta is small enough
to enable a robust r-process and thus lanthanide-rich ejecta \citep[see
  \eg][]{Tanaka2018}.

On the other hand, in the second case, namely, for binaries leading to a
long-lived remnant, we expect that our results could have an impact
especially in those cases where a large disc mass is produced and has
sufficient time to lose a considerable part of in terms of ejected matter
(these binaries, we recall, are those with small mass ratio and rapidly
spinning primary). Under these conditions, the continuous neutrino
irradiation, the asymmetries in the disc, as well as further outflows
from the HMNS itself could influence the bulk properties of the kilonova
emission. To capture these processes correctly, expensive high-resolution
long-term simulations of the post-merger HMNS and disc are necessary
\citep{Metzger2014, Perego2014, Fujibayashi2017,
  Fujibayashi2018}. Carrying out such simulations for the configurations
considered here would yield very similar electromagnetic emissions with a
blue component as for systems at lower masses \citep{Wollaeger2017,
  Kawaguchi2018b, Even2020, Korobkin2021, Wollaeger2021}, but in
combination with a potentially unexpectedly high chirp mass, hinting
either to a highly spinning companion, an extreme mass asymmetry or both.

\begin{figure*}
  \centering \centering
 \includegraphics[width=0.9\textwidth]{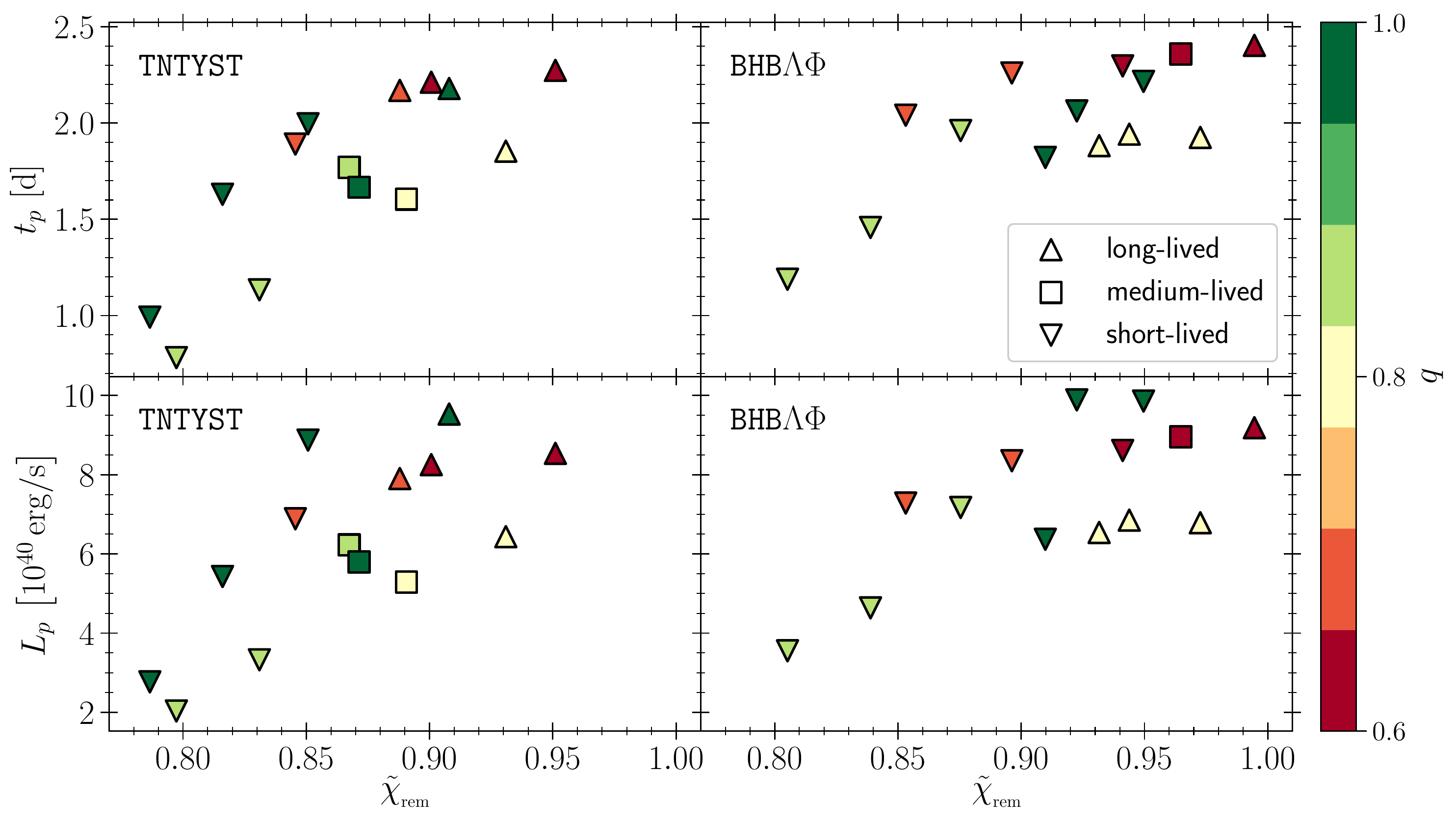}
  \caption{\textit{Left panel:} Behaviour of the characteristic
    properties of the kilonova emission, \ie the time of peak emission
    (top part) and the luminosity at peak (lower part) for binaries
    modeled with the \texttt{TNTYST} EOS and shown as a function of the
    total dimensionless spin at merger of the remnant. Different symbols
    follow the convention presented in Tab.~\ref{tab:initial} and refer
    to the different lifetime of the remnants. Shown with a colourcode is
    the corresponding mass ratio in the binary. \textit{Right panel:} the
    same as the left panel but for the \texttt{BHB}$\Lambda\Phi$
    EOSs.}
  \label{fig:kilonova}
\end{figure*}

As a way to estimate the most salient characteristics of a
single-component kilonova signal, a number of semi-analytical models have
been developed by \citet{Grossman2014} and \citet{Wollaeger2017} to
estimate the time of peak emission $t_p$, the corresponding peak
bolometric luminosity $L_p$, and its effective temperature $T_p$. We here
assume a grey opacity of $10 \, \rm{cm^2 / g}$, which leads to a good
agreement of the peak characteristics for ejecta undergoing a thorough
r-process computed with more sophisticated modelling
\citep{Wollaeger2017}. Furthermore, lacking a robust and universal
estimate, we consider as a reference value that $50\%$ of the disc
rest-mass is ejected and at an average velocity of $\langle v
\rangle_{\rm disc} = 0.1$. We recall that the total ejected mass $M_{\rm
  tot}$ is the sum of the dynamical and disc ejecta [\cf
  Eq.~\eqref{eq:Mtot}] and we use these two distinct masses obtain
mass-weighted average velocity
\begin{align}
\langle v \rangle := \frac{\langle v \rangle_{\rm ej}\, M_{\rm ej} + \tfrac{1}{2}\langle v
  \rangle_{\rm disc}\, M_{\rm disc}}{M_{\rm tot}} \,,
\end{align}
The resulting characteristics of the kilonova emission are listed in
Tab.~\ref{tab:ejecta-kn}, which includes long-lived remnants (indicated
between round brackets) and for which the unknown lifetime of the remnant
can potentially have large impact on the composition of the secular
ejecta leading to more intricate emission features \citep{Wollaeger2017,
  Even2020, Korobkin2021}. At the same time, Fig.~\ref{fig:kilonova}
reports, as a function of the total dimensionless spin of the merger
remnant, the time of the peak (top part) and the luminosity at peak
emission (lower part) for binaries modeled either with the \texttt{TNTYST}
EOS (left panel) or with the \texttt{BHB}$\Lambda\Phi$ EOSs (right
panel). Different symbols follow the remnant lifetime convention
presented in Tab.~\ref{tab:initial} and the corresponding mass ratio
in the binary is shown with a colourcode.

Notwithstanding the inherent scattering of the data,
Fig.~\ref{fig:kilonova} clearly shows that there is a nonlinear
correlation between the peak properties of the kilonova emission and the
dimensionless spin of the remnant at merger. More specifically, large
spins will systematically lead to peak times $t_p$ taking place up to one
day later and to luminosities $L_p$ that are up to a factor five
larger. Not surprisingly, these later peaks and larger luminosities
correspond to models that are long-lived and could survive also beyond
the timescale of our simulations. Hence, these values are to be
interpreted only as lower limits both for $t_p$ an $L_p$. Interestingly,
because small mass ratios, \ie $q=0.6$, are in general needed to produce
long-lived remnants, it is such configurations that lead to the largest
peak times and luminosities. Finally, for the \texttt{BHB}$\Lambda\Phi$
EOS, the intermediate mass ratios, \ie $q\approx0.84$, yield the smallest
values of $t_p$ and $L_p$; such a clear behaviour is not found in the
\texttt{TNTYST} EOS, for which also equal mass binaries, $q=1$, can be
similarly less luminous and peak early.

Overall, and at least for this simplified modelling, the main source of
variation in determining the characteristics properties of the kilonova
emission is represented by the total ejected mass. Further differences
could emerge from long-lived remnants, though the lifetime of the
surviving merger remnants is unknown and hence no definitive conclusions
can be found for these systems. Whereas these simple estimates show that
there is an appreciable variation in the kilonova emission as a result of
extreme spins and mass ratios, the maximum relative difference for
binaries with the same total mass is well within the uncertainties of
kilonova modelling, \eg concerning opacities and especially morphology of
the ejecta \citep{Wollaeger2017, Even2020, Korobkin2021}.
This clearly calls for improved modelling of the actual kilonova emission.

\section{Discussion}
\label{sec:discussion}

We have investigated the effect of a highly spinning primary companion on
the merger dynamics of high-mass BNS, being the physically more relevant
scenario involving high spin states, where the primary NS was spun-up
throughout the preceding binary evolution by accretion processes. For
these we concentrated on configurations with total masses of up to
$\simeq 11\%$ above their irrotational threshold mass for two fully
temperature-dependent nuclear equations of state, \ie \texttt{TNTYST} and
\texttt{BHB}$\Lambda\Phi$, being in the range of radii and maximum masses
compatible with GW170817 \citep{Most2018,Nathanail2021}. In this way, we
were able to cover for the first time both the space of parameters
ranging from equal-mass to highly asymmetric binaries with mass ratio
$q=0.6$, and the space of parameters of rapidly spinning binaries with
spin aligned with the orbital angular momentum, and where the primary has
dimensionless spin ranging from $\chi_{_{1}} = 0.3$ to $\chi_{_{1}} =
0.6$.

By performing fully general-relativistic hydrodynamic simulations of the
inspiral, merger and post-merger of these systems, we were able to
highlight a number of interesting aspects of the dynamics of these
binaries. In particular, attention has been paid to the development of
asymmetries in the mass distribution of the merger remnant. Besides the
standard $\left(2{,}2\right)$ bar-mode deformation, many remnants have
exhibited also an $\left(2{,}1\right)$ asymmetry, which is an effective
indicator of the development of an $m=1$ instability. In particular, such
an instability has been found to provide sizeable contributions to the GW
spectrum for all of the unequal-mass binaries, although these
contributions remain subdominant (by one or more orders of magnitude)
with respect to the $\left(2{,}2\right)$ bar-mode deformation. Furthermore,
depending on the EOS, the largest $\left(2{,}1\right)$ asymmetry can be
reached for the smallest mass ratio ($q=0.6$) in the case of the softer
\texttt{TNTYST} EOS, or for an intermediate one ($q=0.8$) for the
\texttt{BHB}$\Lambda\Phi$ EOS. Notwithstanding its smaller contribution
on the timescales over which the simulations have been carried out,
the $\left(2{,}1\right)$ deformation appears to be less affected by
dissipation, so that the corresponding emission from the $m=1$ GW
emission could provide sizeable, if not comparable contributions for those
binaries with large mass asymmetry and whose remnants can be long-lived.

Our simulations have also revealed a number of interesting aspects of the
properties of the merger remnant, of its lifetime, of the mass in its
disc, and of the ejected mass. All of these aspects can be briefly
summarised as follows.

\begin{itemize}

\item The spin the primary can have a significant effect on the lifetime
  of the remnant even in the case of binaries with masses that
  significantly supercritical, \ie with $M_{\infty} \simeq 1.1
  M^{^{1,0}}_{\rm th}$. In particular, a sufficiently large dimensionless
  angular momentum at merger is able to systematically yield remnants
  that are long-lived.

\item Binaries with significant mass asymmetry with $q\simeq0.6-0.8$,
  tend to systematically yield remnants that are longer-lived than those
  resulting from equal-mass systems. At the same time, the dependence of
  the remnant lifetime on the mass ratio can either be monotonic, as in
  the case of the softer \texttt{TNTYST}, or show a local maximum for
  intermediate mass ratios, \eg $q=0.8$ for the stiffer
  \texttt{BHB}$\Lambda\Phi$ EOS.

\item The remnant-disc masses, and therefore also the secular ejecta, can
  vary significantly and up to three orders of magnitude in the space of
  parameters considered. In particular, the largest disc masses are
  produced by binaries with large primary spin and small mass ratio; by
  contrast, the smallest disc masses are produced by binaries with small
  primary spin and equal masses.

\item Also the dynamically ejected matter can vary significantly across
  our simulated binaries. More specifically, large (small) amounts of
  ejected mass are again found for binaries with large (small) primary
  spin and small (large) mass ratios. However, the largest amounts of
  dynamical ejecta has been found for binaries with very large primary
  spin but equal masses.

\item A nonlinear correlation is present between the peak properties of
  the kilonova emission and the dimensionless spin of the remnant at
  merger. In particular, large spins will systematically lead to peak
  times $t_p$ delayed of up to one day and to luminosities $L_p$ that are
  up to a factor five larger. While these delays and increased
  luminosities can be interpreted as due to highly spinning binaries or
  small mass ratios, a more accurate modelling of the kilonova emission is
  needed.

\end{itemize}

Overall, the results obtained show that the merger of BNSs with small
mass ratios and rapidly spinning primaries offer exciting prospects for
the detection of an electromagnetic counterpart as they systematically
increase the lifetime of the remnant and boost the ejected mass, either
dynamically or via a secular emission from the disc. It is presently
unclear whether systems with these properties occur in nature, just as it
is difficult from the inspiral signal alone to determine with precision what
is the mass ratio and the spins of the components. This study thus
highlights that a number of hints on binaries with extreme spins and mass
ratios can be found in their electromagnetic counterparts, which we will
model in future work.

\section*{Acknowledgements}
The authors gratefully acknowledge funding by the State of Hesse within
the Research Cluster ELEMENTS (Project ID 500/10.006), by the ERC
Advanced Grant ``JETSET: Launching, propagation and emission of
relativistic jets from binary mergers and across mass scales'' (Grant
No. 884631), and by HGS-HIRe for FAIR. ERM acknowledges support from a
joint fellowship at the Princeton Center for Theoretical Science, the
Princeton Gravity Initiative and the Institute for Advanced Study. Part
of the simulations were performed on the national supercomputer HPE
Apollo Hawk at the High Performance Computing Center Stuttgart (HLRS)
under allocations BBHDISKS and BNSMIC, and the GCS Supercomputer SuperMUC
at Leibniz Supercomputing Centre (www.lrz.de) This work benefited from
the valuable implementations in the kuibit \citep{kuibit21}, SciPy
\citep{SciPy2020}, NumPy \citep{Harris2020} and matplotlib
\citep{Hunter2007} libraries.

\section*{Data Availability}
Data is available upon reasonable request from the Corresponding Author.



\bibliographystyle{mnras}
\bibliography{spinning_collapse} 


\appendix

\section{Measuring the uncertainty in our estimates}
\label{appendix_A}

  With the goal of measuring the uncertainty in the various
  quantities presented in this paper, we have repeated our analysis for
  all of the models at the lower resolution of $h = 0.2 M_{\odot}$ and
  thus a factor in volume resolution of $\sim 2$. We have also carried
  out four simulations at a higher resolution of $h = 0.133 M_{\odot}$,
  which are obviously considerably more expensive. These models are
  selected to be at the high-mass end of the configurations, three with
  the \texttt{TNTYST} EOS and different mass ratios and spins, as well as
  one binary with the \texttt{BHB}$\Lambda\Phi$ EOS. While this analysis
  does not allow us to conduct a detailed convergence analysis, it gives
  us an estimate and approximate upper bound on the resolution-dependent
  uncertainties. These uncertainties have been collected in four distinct
  classes depending on the quantity considered.  They are: a) inspiral
  and post-merger GW radiation; b) post-merger survival time of the HMNS;
  c) dynamical ejecta; d) disc mass and thus the approximate total
  ejected mass on secular time-scales. All of these uncertainties are
  discussed in detail below. In summary, our uncertainties are of few
  tens of percent at most; in this sense, they are comparable or smaller
  than those reported by other groups in similar investigations
  \citep{Radice2018a,Nedora2019}.

\begin{itemize}

\item[a)]To assess the discretization-dependent uncertainties in
  the post-merger GW radiation, i.e., including the cumulative effect of
  the quantities presented in Fig.~\ref{fig:GW_quantities}, we calculated
  the relative differences between both resolutions for both the total
  radiated angular momentum and the radiated energy.  In all cases and
  quantities the average absolute relative difference is $12 \%$ (median
  $11 \%$), including a few showing differences of up to $\lesssim 50
  \%$, while no correlation with the absolute magnitude of the particular
  quantity is detectable. Additionally, the cumulative post-merger
  radiation is on average smaller at lower resolution for both quantities
  and modes.  The inspiral GW emission is completely dominated by the
  $(2,2)$ emission leading to an overall average uncertainty of at most
  $2 \%$ for both quantities. Comparing to the high-resolution (low), the
  maximum relative differences in the post-merger radiation is $\simeq 10
  \%$ ($\simeq 21 \%$), while the averages decrease to $\lesssim4 \%$
  ($\lesssim6 \%$) for the post-merger and $1 \%$ ($2 \%$) for the
  inspiral emission.
  
\item[b)] In the case of the post-merger survival time of the
  HMNS, most of the short lifetimes are well below the low-resolution
  sampling rate, while one of these models reaches a relative difference
  in the survival time of $50 \%$. At larger survival times, the same
  reach a difference of $30 \%$. Overall, discretization-dependent
  uncertainty in the estimate of the lifetime is of $\lesssim 8 \%$
  (median $\lesssim 3 \%$). A correlation with increasing survival times
  is expected, but due to the sparsity of intermediate lifetimes, this
  could not be verified. Nonetheless, the survival times are on average
  smaller at lower resolution. For the high-resolution (low-) sample,
  with three collapsing models, the relative difference in the lifetime
  estimate is $\simeq1 \%$ ($3 \%$). It should be noted though that, due
  to the high costs of these simulations, we do not include models with
  intermediate or longer lifetimes (the lifetime is $1.17 \, \rm{ms}$ at
  most within this sample), so that significantly larger relative
  differences can develop over longer lifetimes also at these rather high
  resolution.

\item[c)] For the dynamical ejecta we employed two different
  uncertainty estimates. First, again, the differences between the total
  ejected mass is compared between different resolutions. In addition to
  the standard-resolution result, we also compute the relative
  differences with respect to the extrapolated total ejected mass $M_{\rm
    tot}$. The binaries with the smallest dynamically ejected mass show
  the largest deviations between the two resolutions, reaching $\lesssim
  150 \%$. This is because of the very low densities reached in these
  cases that can be masked into the atmosphere in the low-resolutions
  outer regions of the simulation domain. Note that these differences are
  nontheless insignificant when considering the total mass ejected, whose
  relative difference is still well below $0.1 \%$. Overall, the average
  absolute relative differences in the measurement of the dynamically
  ejected mass is $26 \%$ (median $20 \%$).  We have also explored how
  our estimates change with the placement of the outer detectors. In
  particular, we have recomputed the dynamically ejected masses at a
  distance of $443\,{\rm km}$ from the merger remnant and compared the
  result with those measured at $295\,{\rm km}$. In this case, we see a
  weak correlation with the total ejected mass, where the differences
  peak at $28 \%$ together with an average of $7 \%$ (median $3 \%$). The
  observed correlation confirms the validity of our choice in a detector
  that is closer to the merger remnant. This is because the resolution
  drops outwards in the simulation domain -- due to the box-in-box mesh
  refinement -- so that the mass estimates are less accurate especially
  when lower densities towards atmosphere levels are involved. The
  high-resolution sample of models includes the two cases with strongly
  suppressed dynamical ejecta leading to large uncertainties of $100 \%$
  and $163 \%$, respectively. This indicates that the resolution in the
  outer regions of the grid is still too coarse to draw conclusions that
  are accurate beyond an order of magnitude estimate. When considering
  the whole sample of high-resolution (low-) binaries, the average
  relative uncertainty decreases to $65 \%$ ($80 \%$). This is because
  the other two binaries have far smaller relative uncertainties of $1
  \%$ ($5-20 \%$).

\item[d)] Important for the electromagnetic follow-up emission
  are the uncertainties in the disc mass and thus the mass fraction
  ejected on secular timescales [see Eq.~\eqref{eq:Mtot}]. As in the case
  of the dynamical ejecta, we examined two independent sources of
  uncertainty. Due to lack of full three-dimensional datasets needed to
  measure the discs around the HMNS, we limit our analysis to the models
  collapsing within the simulation time. For these cases, the
  discretization-dependent uncertainty estimate reaches a maximum value
  of $76 \%$, but the deviation is much smaller for the most massive
  discs, while the average is $\approx 10 \%$ (median $3 \%$).  Within
  the high resolution sample we included the largest deviating model
  metioned above. The outlier drops from $76 \%$ to mere $1 \%$
  indicating that the low-resolution simulation is not able to capture
  the full dynamics involving the creation of the disc accurately. At the
  same time, also the other uncertainties decrease leading to an average
  and median of the high-resolution (low-) sample of both $2 \%$ (average
  $28 \%$, median $7 \%$). In summary, the discretization error on the
  extrapolated total ejected mass $M_{\rm tot}$ is most probably
  dominated by the disc-mass estimates, except for the models with the
  largest dynamical ejecta. \\ A second source of error, which inevitably
  leads to an overestimation of the disc mass, is that it is not possible
  to reach a perfectly stationary disc over the timescales
  simulated. Since the disc mass measurements reach quasi-constant values
  on exponential timescales, we model the future accreted matter by an
  exponentially decaying contribution to the disc mass estimate. This is
  underpinned by the fact that the post-collapse tail of the accretion
  rate, i.e., the time derivative of the disc mass estimate, is observed
  to be log-linear \citet{Rezzolla:2010}. Fitting this tail gives us an
  estimate on the yet to be accreted mass for $t \to \infty$.  The
  maximal value reaches $36 \%$, with an average of $14 \%$ (median $13
  \%$). While these represent a significant systematic effect with
  respect to the true post-collapse disc mass, we do not correct for them
  here.  We note that the evolution and subsequent effect of magnetic
  fields and neutrino absorption are important for the equilibration of
  the disc, which are not included in our simple estimates of the
  long-term accretion.

\end{itemize}


\bsp	
\label{lastpage}
\end{document}